\documentclass[twocolumn, twocolappendix]{aastex63}

\usepackage{savesym}
\savesymbol{tablenum}
\usepackage{bm, siunitx}
\usepackage{lineno,multirow,booktabs}
\received{December 22, 2020}
\revised{July 11, 2021}
\accepted{August 1, 2021}
\submitjournal{The Astrophysical Journal Supplement Series}
\shorttitle{Properties of BL Lacs at low frequencies}
\shortauthors{Mooney et al.}

\begin{document}

\title{Characterising the extended morphologies of BL Lacs at 144 MHz with LOFAR}

\correspondingauthor{Se\'an Mooney}
\email{sean.mooney@ucdconnect.ie}
\author[0000-0003-0225-7355]{Se\'an Mooney}
\affiliation{School of Physics, University College Dublin, Belfield, Dublin 4, Republic of Ireland}
\author[0000-0002-1704-9850]{Francesco Massaro}
\affiliation{Dipartimento di Fisica, Universit\'a degli Studi di Torino, via Pietro Giuria 1, 10125 Torino, Italy}
\author[0000-0002-4855-2694]{John Quinn}
\affiliation{School of Physics, University College Dublin, Belfield, Dublin 4, Republic of Ireland}
\author[0000-0003-3684-4275]{Alessandro Capetti}
\affiliation{INAF - Osservatorio Astrofisico di Torino, via Osservatorio 20, Pino Torinese, 10025 Italy}
\author[0000-0002-1824-0411]{Ranieri D. Baldi}
\affiliation{Istituto di Radioastronomia - INAF, Via P. Gobetti 101, I-40129 Bologna, Italy}
\affiliation{School of Physics and Astronomy, University of Southampton, Southampton, SO17 1BJ, UK}
\author[0000-0002-9777-1762]{G\"ulay G\"urkan}
\affiliation{Th\"uringer Landessternwarte, Sternwarte 5, D-07778 Tautenburg, Germany}
\author[0000-0003-4223-1117]{Martin J. Hardcastle}
\affiliation{Centre for Astrophysics Research, School of Physics, Astronomy \& Mathematics, University of Hertfordshire, College Lane, Hatfield AL10 9AB, United Kingdom}
\author[0000-0002-3533-8584]{Cathy Horellou}
\affiliation{Chalmers University of Technology, Department of Space, Earth and Environment, Onsala Space Observatory, SE-43992 Onsala, Sweden}
\author[0000-0001-5649-938X]{Beatriz Mingo}
\affiliation{School of Physical Sciences, The Open University, Walton Hall, Milton Keynes, MK7 6AA, United Kingdom}
\author[0000-0002-9482-6844]{Raffaella Morganti}
\affiliation{ASTRON, the Netherlands Institute for Radio Astronomy, Postbus 2, 7990 AA Dwingeloo, the Netherlands}
\affiliation{Kapteyn Astronomical Institute, University of Groningen, PO Box 800, 9700 AV Groningen, the Netherlands}
\author[0000-0002-3968-3051]{Shane O'Sullivan}
\affiliation{School of Physical Sciences \& CfAR, Dublin City University, Glasnevin, Dublin 9, Republic of Ireland}
\author[0000-0002-7306-1790]{Urszula Pajdosz-\'Smierciak}
\affiliation{Astronomical Observatory of the Jagiellonian University, Orla 171, 30-244 Cracow, Poland}
\author[0000-0001-5829-1099]{Mamta Pandey-Pommier}
\affiliation{USN, Station de Radioastronomie de Nan\c{c}ay Observatoire de Paris route de Souesmes 18330 Nan\c{c}ay France \& Univ Lyon, Univ Lyon1, Ens de Lyon, CNRS, Centre de Recherche Astrophysique de Lyon UMR5574, 9 av Charles Andr\'{e} F- 69230, Saint-Genis-Laval, France}
\author[0000-0001-8887-2257]{Huub R\"ottgering}
\affiliation{Leiden Observatory, Leiden University, PO Box 9513, NL-2300 RA Leiden, the Netherlands}

\begin{abstract}
We present a morphological and spectral study of a sample of $99$ BL Lacs using the LOFAR Two-Metre Sky Survey Second Data Release (LDR2).
Extended emission has been identified at \si{GHz} frequencies around BL Lacs, but with LDR2 it is now possible to systematically study their morphologies at \SI{144}{MHz}, where more diffuse emission is expected. %
LDR2 reveals the presence of extended radio structures around $66/99$ of the BL Lac nuclei, with angular extents ranging up to \SI{115}{\arcsecond}, corresponding to spatial extents of \SI{410}{kpc}.
The extended emission is likely to be both unbeamed diffuse emission and beamed emission associated with relativistic bulk motion in jets.
The spatial extents and luminosities of the extended emission are consistent with the AGN unification scheme where BL Lacs correspond to low-excitation radio galaxies with the jet axis aligned along the line-of-sight.
While extended emission is detected around the majority of BL Lacs, the median \SIrange{144}{1400}{MHz} spectral index and core dominance at \SI{144}{MHz} indicate that the core component contributes \SI{\sim42}{\percent} on average to the total low-frequency flux density.  %
A stronger correlation was found between the \SI{144}{MHz} core flux density and the $\gamma$-ray photon flux ($r = 0.69$) compared to the \SI{144}{MHz} extended flux density and the $\gamma$-ray photon flux ($r = 0.42$). %
This suggests that the radio-to-$\gamma$-ray connection weakens at low radio frequencies because the population of particles that give rise to the $\gamma$-ray flux are distinct from the electrons producing the diffuse synchrotron emission associated with spatially-extended features.
\end{abstract}


\section{Introduction} \label{sec:introduction}

Blazars, interpreted to be radio-loud active galactic nuclei (AGN) with relativistic jets oriented at small angles along the line of sight, are the largest known population of sources in the extragalactic $\gamma$-ray sky.
The small inclination angles result in strong Doppler beaming effects that significantly boost the observed flux of the jet moving towards the Earth and deboost the flux of the counter jet. The jets eventually dissipate by either decollimation or by depositing their kinetic energy in a terminal shock. These processes produce spatially-extended diffuse unbeamed emission which is potentially observable at low frequencies in addition to the beamed emission \citep[as indicated by simulations in, e.g.,][]{2014MNRAS.443.1482H,2016A&A...596A..12M}.

Blazars are classified as either BL Lacs, which typically exhibit narrow emission or absorption lines in their optical spectra, or flat-spectrum radio quasars, which have broad emission lines typical of quasars \citep{2001MNRAS.323..757L}.
Hereinafter we adopt the nomenclature of the blazar catalogue Roma-BZCAT \citep{2015Ap&SS.357...75M}, labelling the former class of BL Lacs as BZBs and the latter as BZQs.
BZBs are the focus of this study their emission is more dominated by non-thermal components when compared to BZQs, which can have features such as a dusty torus and a big blue bump in their broad-band spectral energy distribution \citep{2004ASPC..311...37W}.

According to the unification scheme of radio-loud AGN \citep{1995PASP..107..803U}, source orientation with respect to the line-of-sight is key to explaining observational differences between blazars and radio galaxies, with low-excitation and high-excitation radio galaxies \citep[LERGs and HERGs;][]{1979MNRAS.188..111H} believed to be the parent populations of BZBs and BZQs, respectively. A strong prediction of the unification theory is that at low frequencies the morphologies of BZBs should conform with LERGs viewed at small angles to the jet \citep[see the review by][]{2012MNRAS.421.1569B}. For highly aligned sources the low-frequency radio emission is expected to consist of an unresolved beamed core surrounded by extended diffuse emission related to the jet terminus, while for jets with non-zero inclination angles, in addition to the core, there may be emission associated with the large-scale jet and potentially the counterjet including both termination regions.

While there have been GHz studies that detected extended emission  \citep[e.g][]{1983ApJ...266...18U,1985ApJ...294..158A,1993AJ....106..875L}, to date, the BZBs morphologies at $\leq\SI{144}{MHz}$ have not been systematically explored because no wide-field survey has had the spatial resolution, dynamic range, and sensitivity required to resolve all the components of the sources. Information about the beamed and
extended emission has been inferred from spectral studies, where the beamed emission is expected to follow a power law with a flat spectral index (i.e. $-0.5 \leq \alpha \leq 0.5$, where $S_\nu \propto \nu^{\alpha}$ throughout) and the diffuse emission is believed to have a power law spectral index more typical of optically-thin synchrotron emission ($\alpha\approx-0.8$).
The flat-spectrum beamed component dominates the radio spectrum above $\sim\SI{1}{GHz}$ \citep[e.g.][]{2007ApJS..171...61H} and in the last decade it has been shown that the spectral indices of blazars are generally flat below ${\sim}\SI{1}{GHz}$ as well \citep{2013ApJS..208...15M}.
Indeed, this characteristic flat spectrum at low frequencies has been successfully leveraged in identifying $\gamma$-ray sources \citep{2014ApJS..213....3M}.

Blazars tend to be core-dominated in the GHz regime, where the core dominance is a proxy for the beaming factor and hence the inclination angle \citep{1985ApJ...294..158A,1993ApJ...406..430P}. %
In the \SI{\sim100}{MHz} regime, it has been inferred that blazars are core-dominated, on the basis of the flat spectral indices \citep{2014ApJS..213....3M}, but a direct measurement of the core dominances for a sample of blazars has yet to be ascertained. %
\citet{2016A&A...588A.141G} and \citet{2019MNRAS.490.5798D} estimated the low frequency core dominances of blazars using a spectral decomposition technique, relying on assumptions about the spectral indices of the core and extended components respectively. %

Given the broadband nature of blazar spectral energy distributions, a comprehensive understanding of their low frequency spectra is paramount to obtain a complete view of blazars.
This is highlighted by the radio-to-$\gamma$-ray connection, where there is a well-established link between the $\gtrsim\SI{1}{GHz}$ flux and the $\gamma$-ray flux, spanning ${\sim}17$ decades of energy \citep[e.g.][]{2011ApJ...741...30A}. From this, it is inferred that the radio and $\gamma$-ray emission are produced by the same population of relativistic electrons.
The relationship between the flux density at hundreds of MHz and the $\gamma$-ray flux is less clear however \citep{2016A&A...588A.141G,2019A&A...622A..14M}, possibly because the $\gamma$-ray-emitting particles are distinct from the electrons that give rise to the large-scale diffuse emission that is expected to be prevalent at low frequencies.

LOFAR is conducting a high angular resolution, highly-sensitive survey of the northern hemisphere sky at \SI{144}{MHz} \citep[LoTSS; the LOFAR Two-Metre Sky Survey;][]{2017A&A...598A.104S}, with more than \SI{21}{\percent} of the sky observed to date.
The LoTSS Second Data Release (LDR2) presents a unique opportunity to make improved low-frequency radio measurements of blazars and, for the first time, investigate the morphology of blazars at \SI{144}{MHz}.
We present LDR2 data for a sample of BZBs and measure the spatial extents, core dominances, and spectral indices to characterise the diffuse emission.
The radio-to-$\gamma$-ray connection is investigated for the low frequency core and extended components separately in order to understand why this trend, which is clear at GHz frequencies, tends not to be significant at \SI{144}{MHz}.

This paper is organised as follows: In \S~\ref{sec:sample-selection} the sample selection is outlined, in \S~\ref{sec:data-analysis} the analysis is detailed, in \S~\ref{sec:results} the results are presented, in \S~\ref{sec:discussion} the findings are discussed, and in \S~\ref{sec:summary-and-conclusions} a summary is provided. A $\Lambda$CDM cosmological model is used in this paper with $h = 0.70$ \citep{2017Natur.551...85A}, $\Omega_{\mathrm{m}} = 0.26$, and $\Omega_{\Lambda} = 0.74$, where $H_{0} = 100 h \, \si{\kilo \metre} \, \si{\per \second} \, \si{\mega pc^{-1}}$ is the Hubble constant.

\section{Datasets and sample selection} \label{sec:sample-selection}

\subsection{LOFAR dataset} \label{sec:data-analysis:lofar-observations}

The Low Frequency Array \citep[LOFAR;][]{2013A&A...556A...2V} is a radio interferometer with stations located throughout Europe.
One goal of the LOFAR Surveys Key Science Project (SKSP) is to map the northern hemisphere sky at \SIrange{120}{168}{\mega\hertz}; this is known as the LOFAR Two Metre Sky Survey \citep[LoTSS;][]{2017A&A...598A.104S}. In this paper we use data from a subset of LoTSS Second Data Release (LDR2).

LDR2 data were processed with the SKSP pipeline\footnote{\url{https://github.com/mhardcastle/ddf-pipeline}} version $2.2$ as described by \citet{2019A&A...622A...1S}.
In the pipeline, direction-dependent calibration was carried out using \textsc{killMS} \citep{2014A&A...566A.127T, 2015MNRAS.449.2668S} and imaging was done using \textsc{DDFacet} \citep{2018A&A...611A..87T}, where a novel self-calibration strategy was employed \citep[\S~5 of][]{2019A&A...622A...1S,2020arXiv201108328T}. With this strategy, extended emission that is undetected in early cycles of self-calibration is less likely to modelled out, and more attention is paid to properly deconvolving that emission. This more complex self-calibration process improved the sensitivity to extended emission with respect to LoTSS DR1 (Tasse et al. in prep.).
Source catalogues were extracted by the pipeline using \textsc{PyBDSF} \citep{2015ascl.soft02007M}.

\begin{figure}
    \plotone{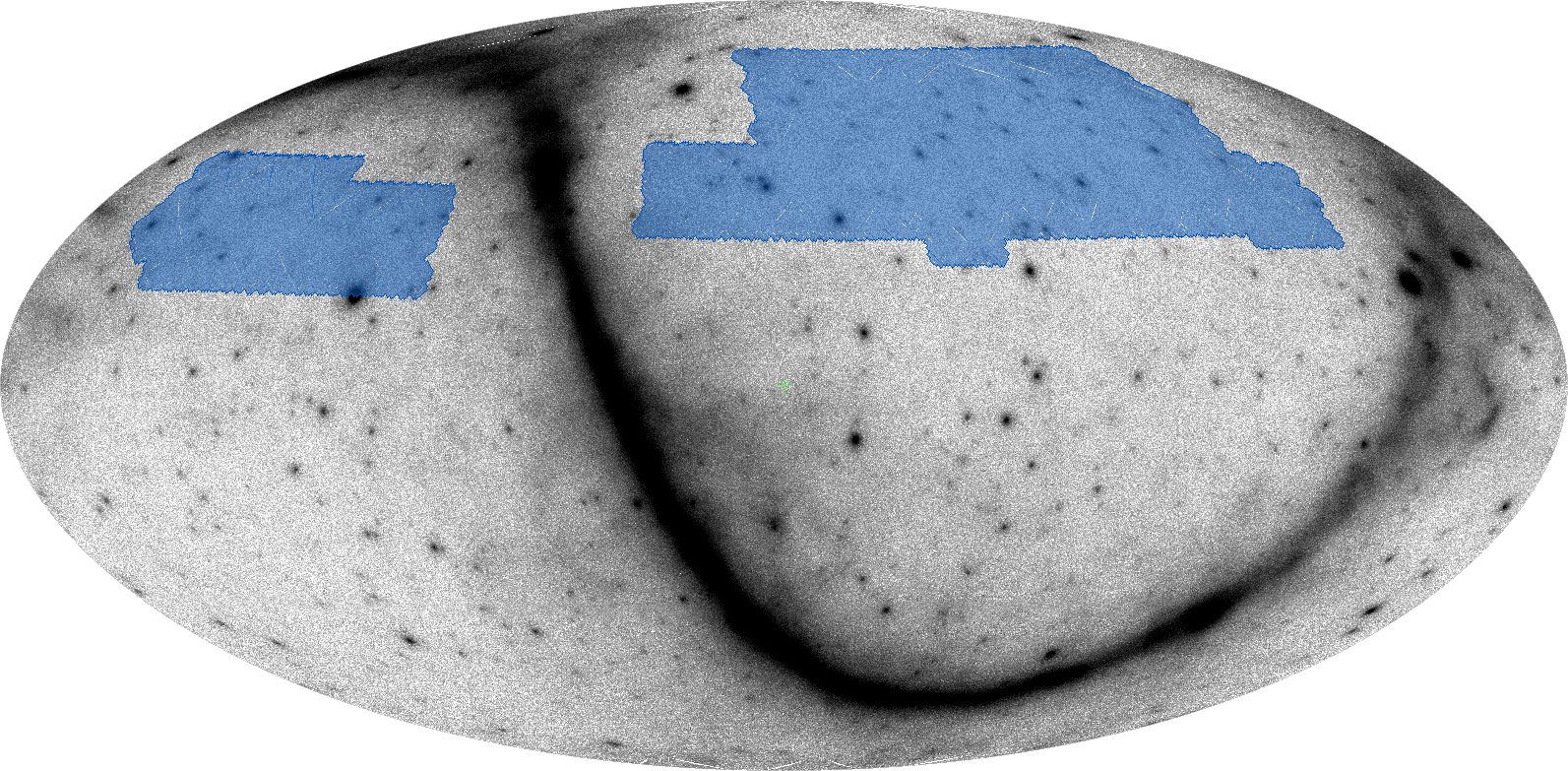}
    \caption{%
    LoTSS DR2 multi-order coverage (blue) overlaid on the Fermi-LAT 0.3–1 GeV HEALPix survey \citep[greyscale;][]{2009ApJ...697.1071A}. Aitoff projection is used. LoTSS DR2 covers 27\% of the northern hemisphere sky as of 2020 February 01. The goal of LoTSS is to survey the entire Northern Hemisphere sky.
    \label{fig:lofarmoc}}
\end{figure}

LDR2 observations are in progress.
We use the LDR2 data that were available as of 2020 February 01, amounting to \SI{4240}{deg^2} of sky coverage, which is \SI{21}{\percent} of the northern hemisphere sky (Fig.~\ref{fig:lofarmoc}).
This LDR2 subset encompasses the publicly-available LDR1 and the catalogue contains \SI{3.6}{M} sources.
The resolution is \SI{6}{\arcsecond} with a pixel size of \SI{1.5}{\arcsecond} and the median RMS noise level is ${\sim}\SI{70}{\micro Jy \, beam^{-1}}$.

\subsection{Ancillary datasets} \label{ancildat}

The Faint Images of the Radio Sky at Twenty-cm \citep[FIRST;][]{1995ApJ...450..559B}\footnote{\url{http://sundog.stsci.edu/}} survey was used, where available, to visually check for the presence of extended emission at \SI{1.4}{GHz}. FIRST data were only unavailable for BL Lacs lying outside its footprint. These data have a comparable resolution to LDR2 (${\sim}\SI{5}{\arcsecond}$).

The \SIrange{144}{1400}{MHz} spectral indices are computed using the NRAO VLA Sky Survey \citep[NVSS;][]{1998AJ....115.1693C}\footnote{\url{https://www.cv.nrao.edu/nvss}} with LDR2. %
NVSS is preferred to FIRST because for extended sources there can be a deficit in the FIRST total flux densities compared to NVSS, particularly for sources with total fluxes in the \SIrange{2}{20}{\milli Jy} range.
This is due to extended flux being undetectable given the lack of short baselines in the FIRST UV coverage \citep{2015ApJ...801...26H}.
However, the FIRST resolution is more favourable than the \SI{45}{\arcsecond} resolution offered by NVSS, so FIRST was used over NVSS to look for large-scale jets at \SI{1.4}{GHz}.

The difference between the spatial resolutions of NVSS and LoTSS are not likely to be problematic \citep{2018MNRAS.474.5008D}. All BL Lacs are mainly point-like in FIRST and NVSS (i.e. consistent with the beam size) and flux densities were computed by integrating over the beam. We also checked the flux densities against those reported in \citet{2016A&A...588A.141G} and \citet{2013ApJS..207....4M} to verify that no flux was missing (where NVSS had similar beam sizes to these surveys).

We used $i$-band data from the Panoramic Survey Telescope and Rapid Response System Data Release 1 \citep[Pan-STARRS1;][]{2016arXiv161205560C}\footnote{\url{https://panstarrs.stsci.edu}} to check that, for each BZB, the emission at detected MHz frequencies is likely associated with its optical counterpart corresponding to the coordinates reported in Roma-BZCAT. Both FIRST and Pan-STARRS1 images for the sample can be accessed through the online appendices.

\subsection{BZB sample selection} \label{sec:sample-selection:bzbs}

We started with the Roma-BZCAT 5\textsuperscript{th} edition \citep{2015Ap&SS.357...75M} blazar catalogue and took the following steps to identify the BZBs that are in LDR2.

\begin{itemize}
    \item We first considered all $3561$ blazars in Roma-BZCAT.
    We added to this $331$ newly identified blazars from a recent optical spectroscopic campaign that will feature in the next release of Roma-BZCAT \citep[][]{2015AJ....149..163L,2015AJ....149..160R,2016AJ....151...95A,2016Ap&SS.361..337M}.
    \item We selected only the BZBs from this enhanced version of Roma-BZCAT that have a counterpart in LDR2 within a few arcseconds, where the criterion depended on the spatial extent of the source. %
    The Roma-BZCAT position, which refers to the central engine, and the LDR2 position, which relates to the centroid of the radio emission, are not necessarily aligned.
    The majority ($55\%$) of BZBs had a crossmatch separation of $<\SI{1}{\arcsecond}$ and $85\%$ of sources had a separation of $<\SI{6}{\arcsecond}$, which is the LDR2 beam width. 
    The largest crossmatch separations pertain to highly extended sources.
    A counterpart was identified for all known BZBs within the LDR2 footprint and all crossmatches were visually confirmed.
    We then excluded BZBs that have uncertain redshift estimates, and $99$ BZBs remained.

    \item We also identified the BZBs in the sample that have a $\gamma$-ray counterpart by crossmatching the Roma-BZCAT positions with the Fourth Catalog of AGN detected by the \textit{Fermi} Large Area Telescope \citep[4LAC;][]{2020ApJS..247...33A,2020ApJ...892..105A} within \SI{5}{\arcsecond}, where we used the position of the associated counterpart in 4LAC. A total of $53$ BZBs are $\gamma$-ray detected in 4LAC.
    \item Using the Roma-BZCAT positions, the BZB sample was cross-matched with FIRST \citep{1995ApJ...450..559B} within \SI{5}{\arcsecond}. There were FIRST counterparts to $83/99$ BZBs. %
    \item Again using the Roma-BZCAT positions, the BZB sample was cross-matched with Pan-STARRS1 \citep{2016arXiv161205560C} within \SI{5}{\arcsecond}, and a counterpart was identified in all cases.
\end{itemize}

The final sample consists of $99$ BZBs; $91$ are from Roma-BZCAT (5\textsuperscript{th} edition) and $8$ are from recent optical spectroscopic campaigns \citep{2019ApJ...871..162P,2019Ap&SS.364...85P,2019A&A...630A..55D,2016Ap&SS.361..337M}. An overview of the selection is given in Table~\ref{tab:source-totals}. %

\begin{deluxetable}{@{}p{4cm}RR@{}}
    \tablecaption{
        Overview of sources in our sample.
        \label{tab:source-totals}
    }
    \tablehead{
        \colhead{} & \colhead{Total} & \colhead{Selected$^{\ast}$} 
    }     
    \startdata
    Roma-BZCAT v$5.0$                                        & $3561$ &      $91$ \\
    Optical campaigns of unassociated $\gamma$-ray sources   &  $331$ &       $8$ \\
    Total                                                    &        &      $99^{\ast\ast}$ \\
    \enddata
    \tablecomments{
        $^{\ast}$BZBs in the LDR2 footprint with a reliable redshift estimate.\\
        $^{\ast\ast}$All $99$ have an optical counterpart in Pan-STARRS, $83$ have a radio counterpart in FIRST, and $53$ have a $\gamma$-ray counterpart in 4LAC.
    }
\end{deluxetable}

For the analysis presented in \S~\ref{sec:data-analysis}, we used the data in the LDR2 catalogue. However, for J1231+3711, the source extraction software (\textsc{PyBDSF}) failed to accurately characterise the widespread diffuse emission that surrounds the compact core. Therefore, instead of using the LDR2 catalogue values for J1231+3711, we reran \textsc{PyBDSF} with a reduced threshold for the flux detection, and used these recalculated values for our analysis.

\section{Data analysis} \label{sec:data-analysis}

The images and the catalogue we constructed based on LDR2 were the starting point of this analysis. The extents, core dominances, and spectral indices of the BZBs were calculated as follows.

\subsection{Angular and spatial extents at 144 MHz} \label{sec:data-analysis:spatial-extents-at-144-mhz}

We used the empirically-derived equation in \S~3.1 of \citet{2019A&A...622A...1S} to determine which BZBs in the sample are resolved (rBZBs) and which BZBs are unresolved (uBZBs).
This equation defines an envelope separating resolved and unresolved sources,
\begin{equation} \label{eq:point-source}
    \mathcal{R} = \frac{S_\mathrm{total}}{S_\mathrm{peak}} - 1.25 - 3.1 \left( \frac{S_\mathrm{peak}}{\sigma}\right)^{-0.53}
\end{equation}
where $S_\mathrm{total}$ is the total flux density of a source, $S_\mathrm{peak}$ is the peak flux, and $\sigma$ is the RMS noise as per the LDR2 catalogue.
If $\mathcal{R} > 0$, then the source is classified as an rBZB, otherwise the source is a uBZB. %
In total, there are $66/99$ rBZBs and $33/99$ uBZBs in the sample.

The intrinsic angular extent, $\Phi$, of each BZB was taken to be the FWHM of the major axis of the source after Gaussian deconvolution of the point spread function (PSF).
For BZBs that consist of multiple Gaussians grouped, the FWHM in the catalogue refers to the 
combination of the major axes of the individual components combined using moment analysis.
The median intrinsic angular extent for the rBZBs is \SI{13(2)}{\arcsecond} and the median upper limit on the intrinsic angular extents for the uBZBs is \SI{4(1)}{\arcsecond}.

We computed the spatial extent, $D$, of each BZB using the intrinsic angular extent and the redshifts, assuming the flat cosmology described in \S~1 \citep{2006PASP..118.1711W}. %
For uBZBs, the angular and spatial extents are reported as upper limits.

\subsection{Core dominances at 144 MHz} \label{sec:data-analysis:core-dominances-at-144-mhz}

The core dominance is the ratio of fluxes from the core and extended regions \citep{1990ApJ...358..159G}, and %
the distribution of relative intensities of the core to the extended emission should reflect the distribution of Doppler factors \citep{1985ApJ...294..158A, 1993ApJ...406..430P}.
Following on from studies such as that of \citet{1997MNRAS.284..541M}, we define the core dominance at \SI{144}{MHz}, $\rho_{144}$, as
\begin{eqnarray} \label{eq:core-dom}
    \rho_{144} &=& \frac{S_\mathrm{core}}{S_\mathrm{ext}}  %
\end{eqnarray}
where $S_\mathrm{core}$ and $S_\mathrm{ext}$ are the core and extended flux densities respectively and the core is defined to be coincident with the AGN central engine. The core dominances are calculated for the rBZBs only.

To calculate $\rho_{144}$ for each rBZB, first the core and extended flux components were separated. 
We simulated a point source that, when convolved with the LDR2 PSF, had the same peak flux as the core in the LDR2 image.
The integration of this Gaussian component represents the core flux density, $S_\mathrm{core}$. We then computed the extended flux density as $S_\mathrm{ext} = S_\mathrm{total} - S_\mathrm{core}$. %
Finally, $S_\mathrm{core}$ and $S_\mathrm{ext}$ were used with Eq.~\ref{eq:core-dom} to yield $\rho_{144}$.

\subsection{Spectral properties at 144 MHz} \label{sec:data-analysis:spectral-properties-at-144-MHz}

As mentioned in \S~\ref{ancildat}, NVSS fluxes were used with LDR2 to compute the %
\SIrange{144}{1400}{MHz} spatially integrated spectral indices of the BZBs. These spectral indices were then used to calculate the total \SI{144}{MHz} specific luminosities of the sources in units of W\,Hz$^{-1}$.

\section{Results} \label{sec:results}

\subsection{Overview}

The median properties of the sample are provided in Table~\ref{tab:summary}, with uncertainties derived from bootstrapping. Values for each BZB are given in Table~\ref{tab:results}, with the full Table accessible with the supplementary material online\footnote{Accessible at \url{http://tiny.cc/bzb-images-mooney-2020} for review.}.

\begin{deluxetable}{@{}lhhRhhRhhR@{}}
    \tablecaption{
        Median properties of the resolved and unresolved BL Lacs. 
        \label{tab:summary}
    }
    \tablehead{
        \colhead{} & & & \colhead{Resolved} & & & \colhead{Unresolved} & & &\colhead{Total} %
    }     
    \startdata
    $N$&$ 34 $ & $ 27 $&$ 68 $&$ 18 $&$ 20 $&$ 31 $&$ 52 $&$ 47 $&$ 99 $ \\
    $z$&$ 0.31 \pm 0.03 $&$ 0.47 \pm 0.03 $&$ 0.38 \pm 0.02 $&$ 0.29 \pm 0.04 $&$ 0.48 \pm 0.03 $&$ 0.37 \pm 0.04 $&$ 0.31 \pm 0.02 $&$ 0.47 \pm 0.02 $&$ 0.38 \pm 0.02 $\\
    $\log_{10} L_{144}$&$  25.7\pm0.1 $&$ 9 \pm 3 $&$25.6\pm0.1 $&$ 3 \pm 2 $&$ 3 \pm 1 $&$ 24.8\pm0.2 $&$ 4 \pm 1 $&$ 6 \pm 2 $&$ 25.3\pm0.2 $\\
    $\alpha^{1400}_{144}$&$ -0.22 \pm 0.07 $&$ -0.37 \pm 0.06 $&$ -0.37 \pm 0.06 $&$ -0.14 \pm 0.05 $&$ -0.2 \pm 0.1 $&$ -0.12 \pm 0.05 $&$ -0.19 \pm 0.05 $&$ -0.31 \pm 0.04 $&$ -0.30 \pm 0.03 $\\
    $\rho_{144}$&$ 1.7 \pm 0.3 $&$ 1.1 \pm 0.3 $&$ 0.74 \pm 0.06 $&\nodata&\nodata&\nodata&\nodata&\nodata&\nodata\\
    $D$ (\si{kpc}) &$ 201 \pm 23 $&$ 234 \pm 25 $&$ 69 \pm 4 $&\nodata&\nodata&\nodata&\nodata&\nodata&\nodata\\
    $\Phi$ (\si{\arcsecond}) &$ 34 \pm 4 $&$ 41 \pm 7 $&$ 13 \pm 2 $&\nodata&\nodata&\nodata&\nodata&\nodata&\nodata\\
    \enddata
    \tablecomments{Number of sources is $N$, redshift is $z$, \SI{144}{MHz} luminosity is $L_{144}$,
    \SIrange{144}{1400}{MHz} spectral index is $\alpha^{1400}_{144}$, \SI{144}{MHz} core dominance is $\rho_{144}$, spatial extent is $D$, and angular extent is $\Phi$.}%
\end{deluxetable}

\subsection{Visual inspection of the BZB images}

LDR2, FIRST and Pan-STARRS1 $i$-band images of all $99$ BZBs are included with the supplementary material online$^2$. Broadly, we identify three classes of morphology for the BZBs. However, several sources have highly complex, asymmetrical morphologies, and in-depth studies using spatially-resolved spectral information would be necessary to fully interpret the source structures. While imaging artefacts cannot be categorically ruled out for all sources, they do not appear to be a significant factor in any case.

Regarding morphology, firstly there are the $33$ BZBs that do not appear to be extended in LDR2; these sources are the uBZBs by definition.

\begin{figure*}
    \gridline{
        \fig{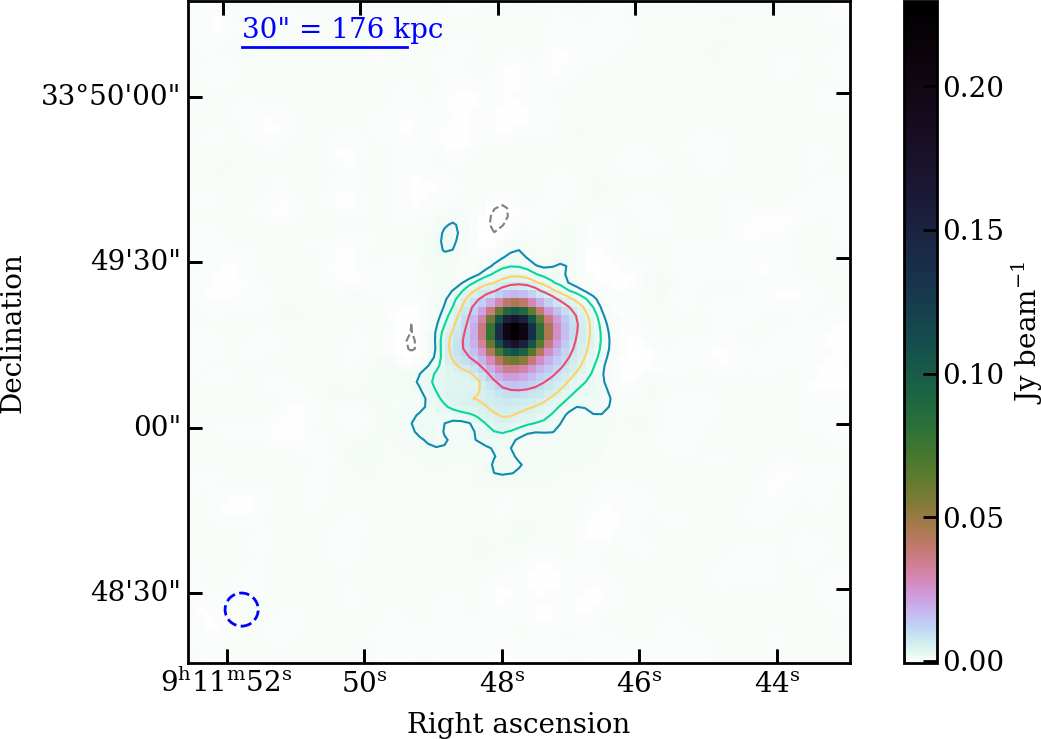}{0.32\textwidth}{(i) J0911+3349 %
        \label{fig:j10911i}}
        \fig{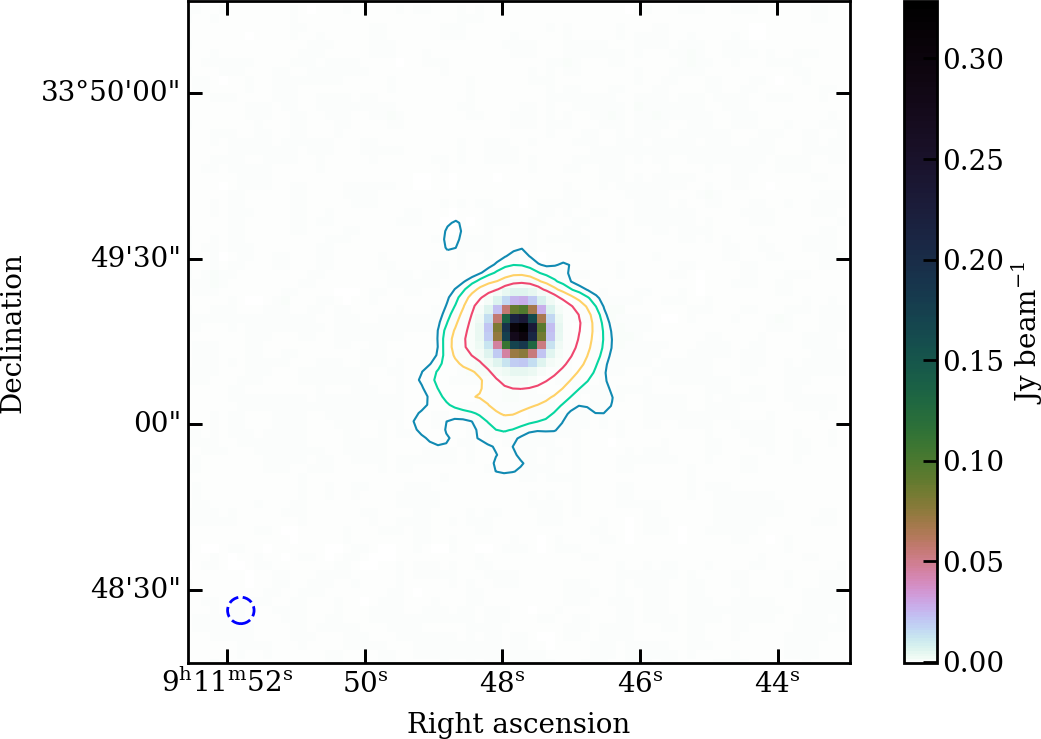}{0.32\textwidth}{(ii) J0911+3349 %
        \label{fig:j10911ii}}
        \fig{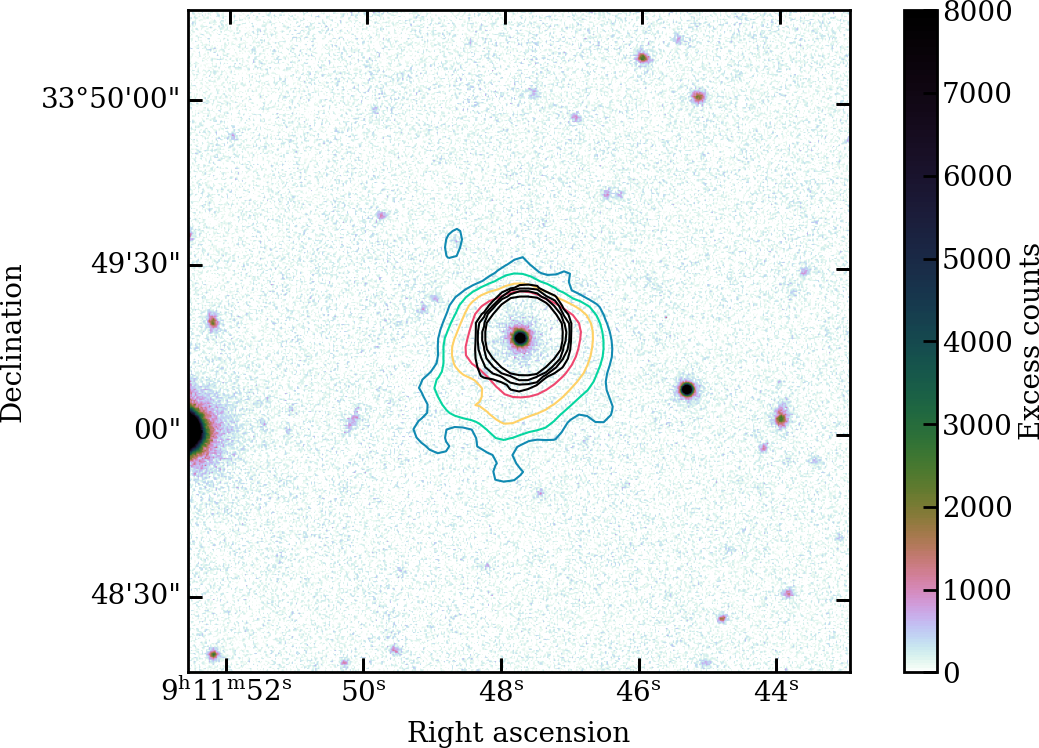}{0.32\textwidth}{(iii) J0911+3349 %
        \label{fig:j10911iii}}
    }
    \gridline{      
        \fig{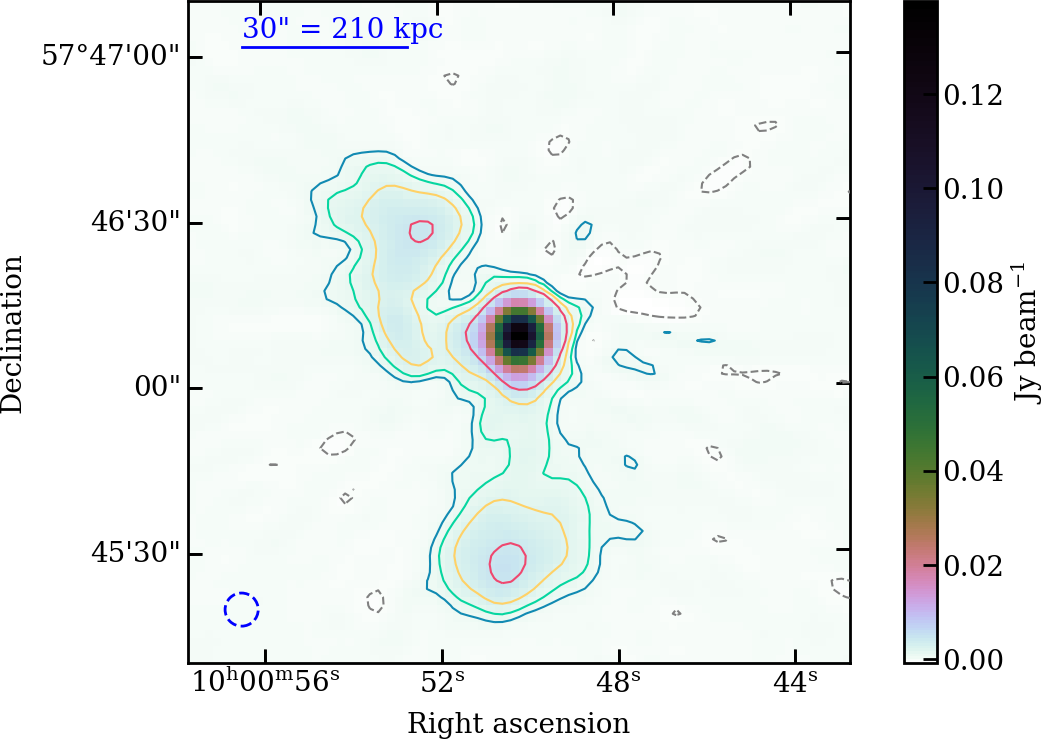}{0.32\textwidth}{(iv) J1000+5746 %
        \label{fig:j1000i}}
        \fig{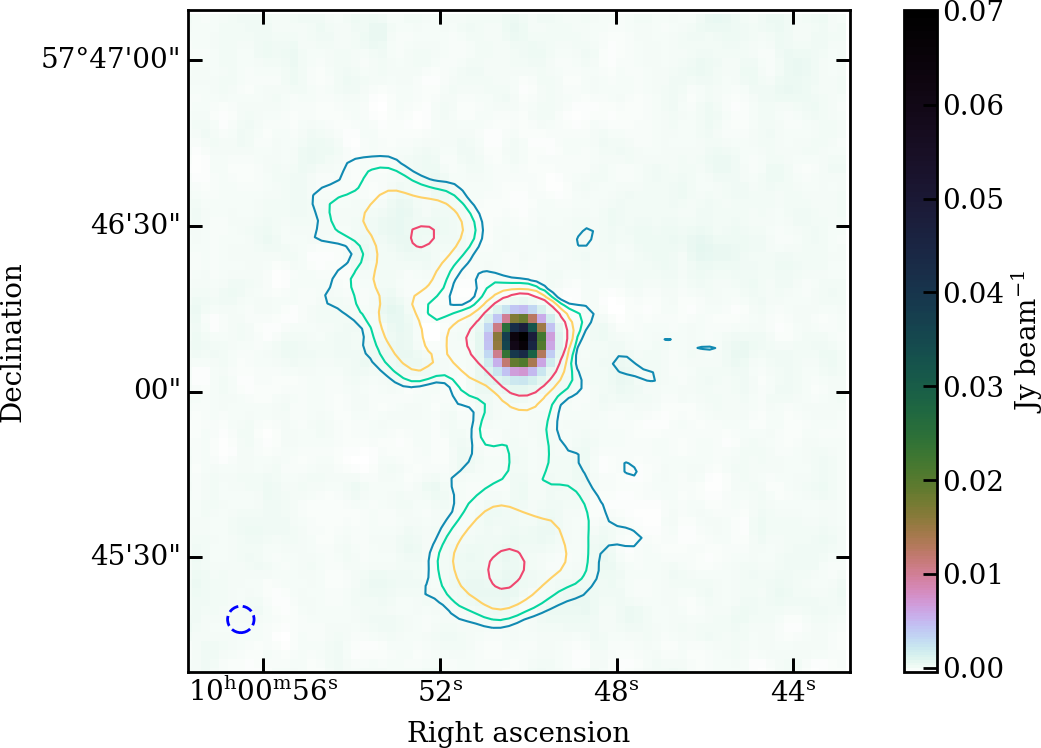}{0.32\textwidth}{(v) J1000+5746 %
        \label{fig:j1000ii}}
        \fig{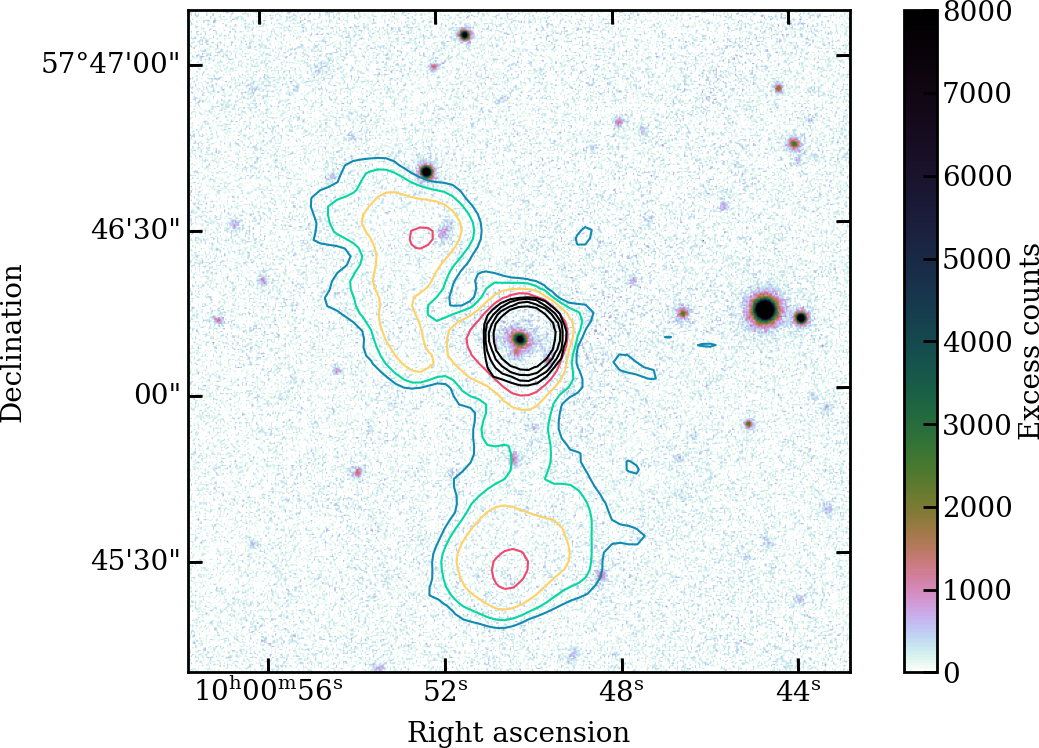}{0.32\textwidth}{(vi) J1000+5746 %
        \label{fig:j1000iii}}
    }
    \gridline{
        \fig{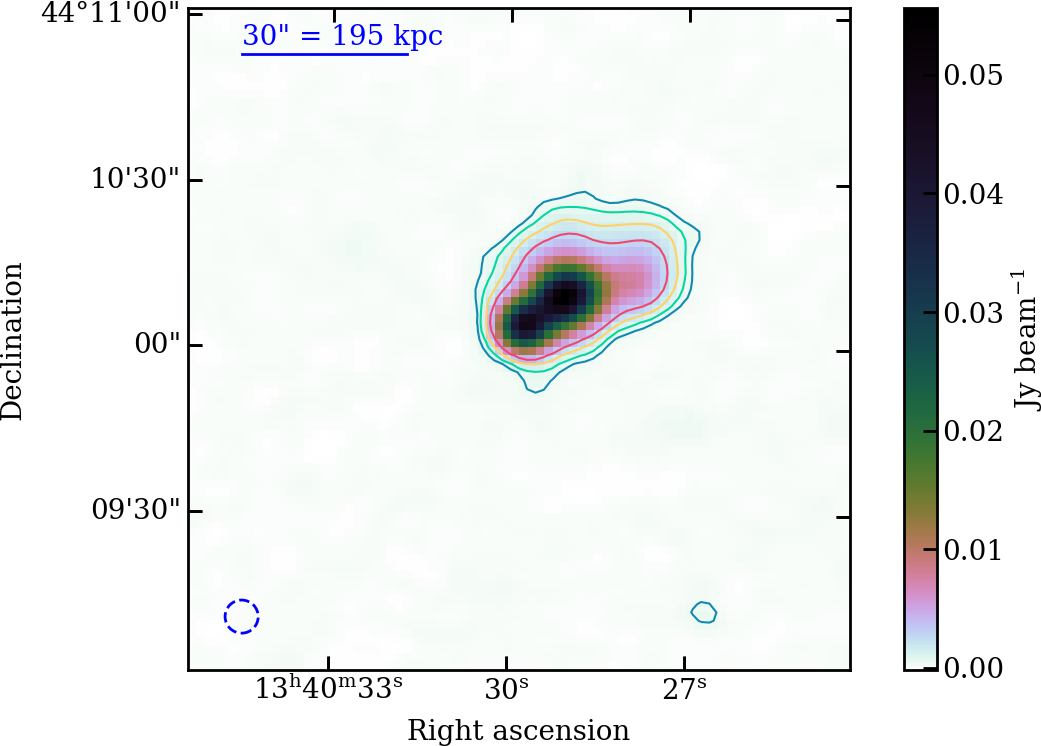}{0.32\textwidth}{(vii) J1340+4410 %
        \label{fig:J1340i}}
        \fig{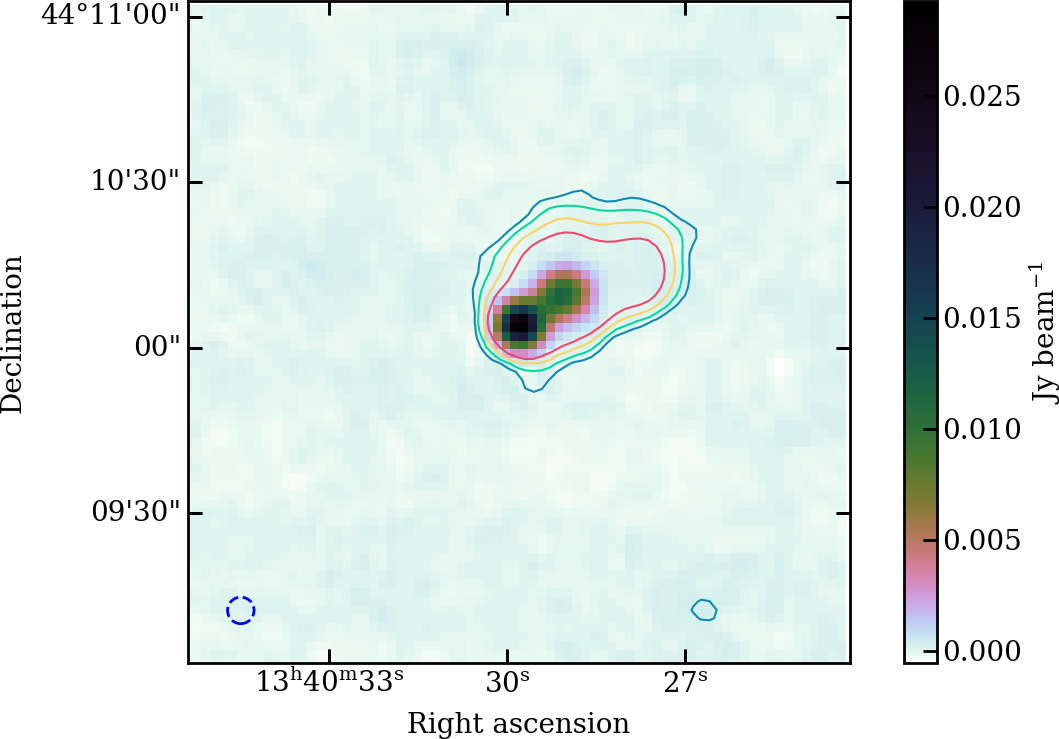}{0.32\textwidth}{(viii) J1340+4410 %
        \label{fig:J1340ii}}
        \fig{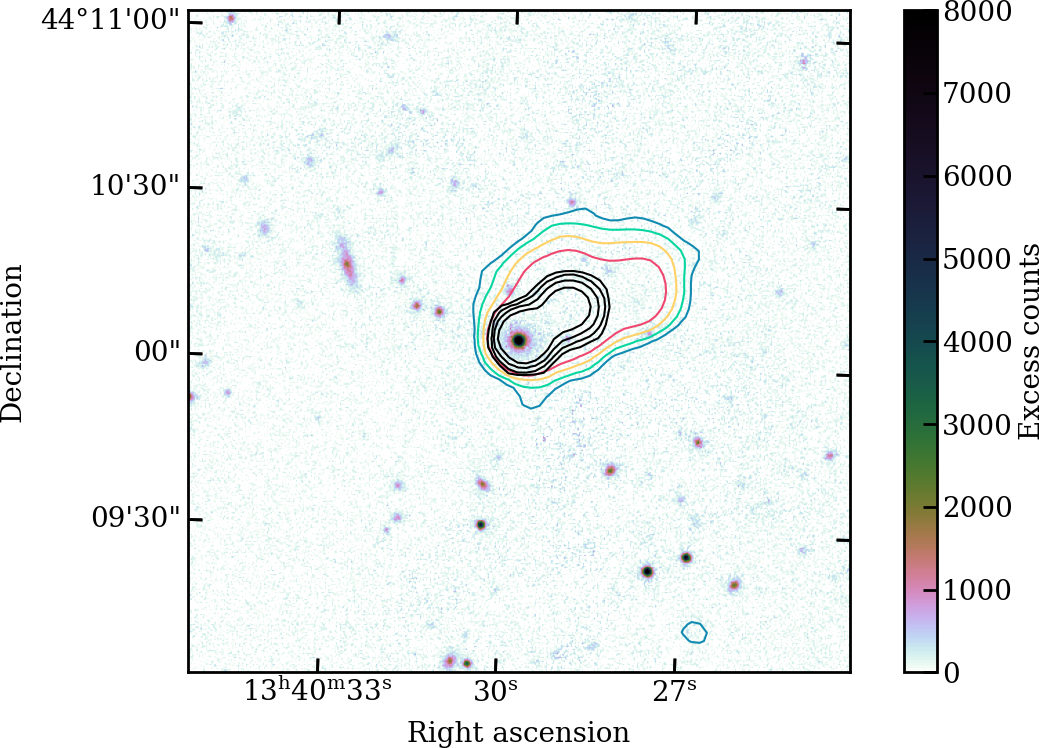}{0.32\textwidth}{(ix) J1340+4410 %
        \label{fig:J1340iii}}
    }
    \caption{Multiwavelength images of J0911+3349 (i--iii), J1000+5746 (iv--vi), and J1340+4410 (vii--ix). %
    The complete figure set (281 images) is available in the online journal.
    (i, iv, vii) LDR2 images with LDR2 contours.
    Contours mark $5\sigma$, $10\sigma$, $20\sigma$, and $40\sigma$ levels at \SI{144}{MHz}, where $\sigma$ is the local RMS noise.
    Dashed grey contours mark the $-3\sigma$ level.
    Colour bars range from \SI{0}{mJy\,beam^{-1}} to the peak flux in each image.
    The dashed blue circle is the FWHM of the LDR2 PSF.
    (ii, v, viii) FIRST maps with LDR2 contours.
    (iii, vi, ix) Pan-STARRS $i$-band images with LDR2 contours.
    \label{fig:joined}}
\end{figure*}

The second group of sources are the rBZBs where the extended morphology seems to consist exclusively of diffuse emission; there are $52/99$ such BZBs. This unbeamed emission is typically of low surface brightness and is believed to be emission from the extended lobes, where the jet interacts with the intergalactic medium. 
These BZBs typically appear as point sources in FIRST because the diffuse emission is not detected. 
When the inclination angle between the jet and our line-of-sight is small, the diffuse emission surrounds the beamed core component (e.g. J0911+3349 in Fig.~\ref{fig:joined}). Alternatively, the inclination angle could be nonzero (but small) or there could be some level of jet bending, in which case there can be a distinct region of diffuse emission separate from the core. For example, for J1000+5746, termination regions associated with the jet and counterjet are potentially observed (Fig.~\ref{fig:joined}).

The remaining $14$ BZBs are also extended, but some fraction of the extended emission is likely to be beamed. These sources are either extended in FIRST or have compact regions of high surface brightness in LDR2, with fluxes comparable to the core (e.g. J1340+4410 in Fig~\ref{fig:joined}). This beamed emission is believed to be associated with relativistic bulk motion in the jet, %
in contrast to the unbeamed emission that is linked with the terminus of the jet. These sources have one-sided jets in FIRST but emission associated with a counterjet is detected in LDR2 in some cases.

\subsection{Angular and spatial extents at 144 MHz} \label{sec:results:spatial-extents-at-144-mhz}

\begin{figure}  %
    \plotone{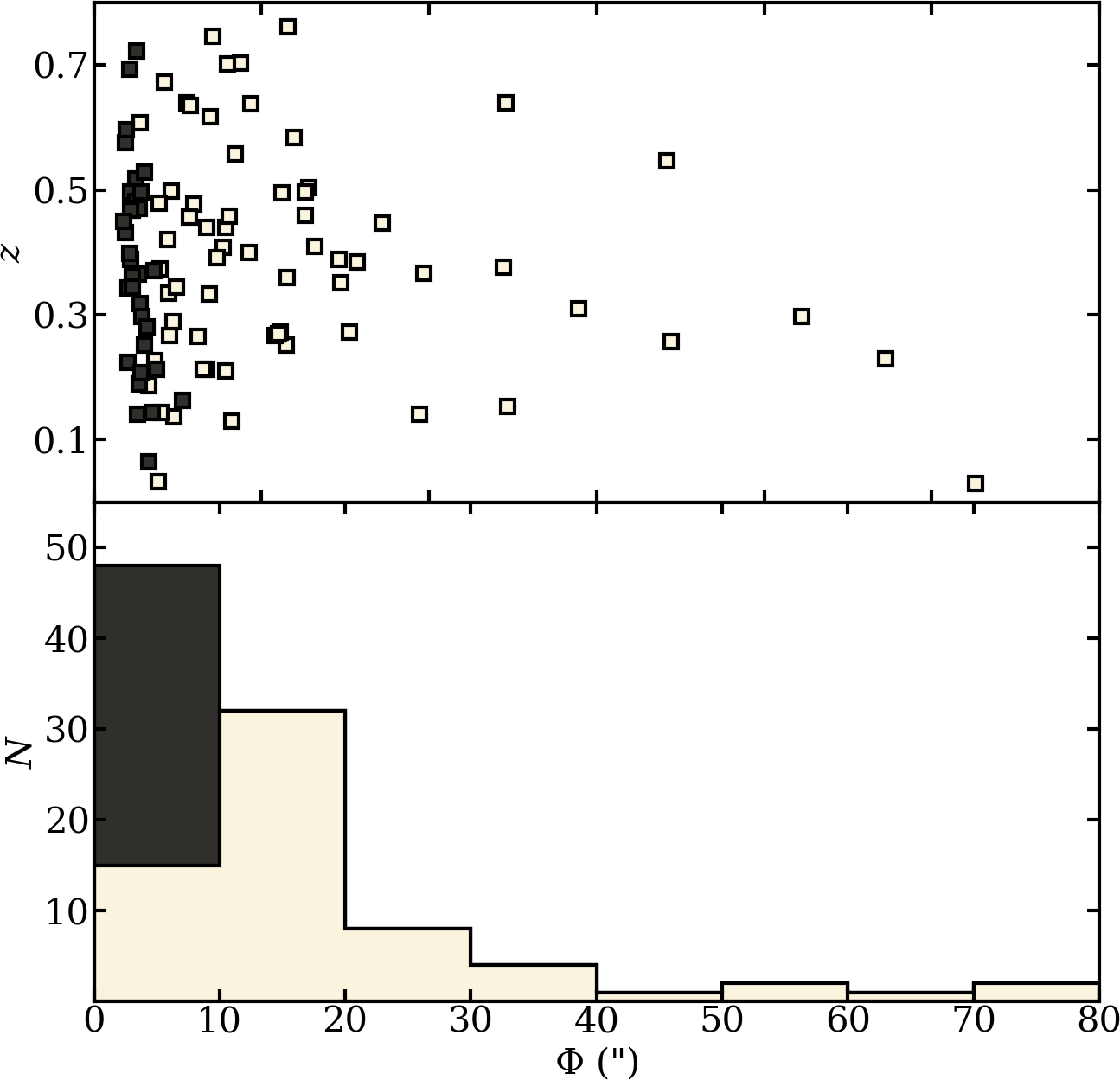}
    \caption{Redshift, $z$, versus intrinsic angular extent, $\Phi$, (top) and the distribution of angular extents (bottom) of the resolved and unresolved BL Lacs.
    \label{fig:plot-angular-extent-redshift}}
\end{figure}

The angular extent distribution is shown in Fig.~\ref{fig:plot-angular-extent-redshift} (bottom) and redshift as a function of angular extent is shown in Fig.~\ref{fig:plot-angular-extent-redshift} (top). The BZBs with the most extended emission tend to be at lower redshifts, but generally there is no trend. The median angular extent of the rBZBs is \SI{13(2)}{\arcsecond}. Note that we classify the sources as rBZBs or uBZBs only to aid the analysis. That is, the uBZBs and rBZBs likely form a continuous distribution rather than representing distinct BZB subclasses. 

\begin{figure}
    \plotone{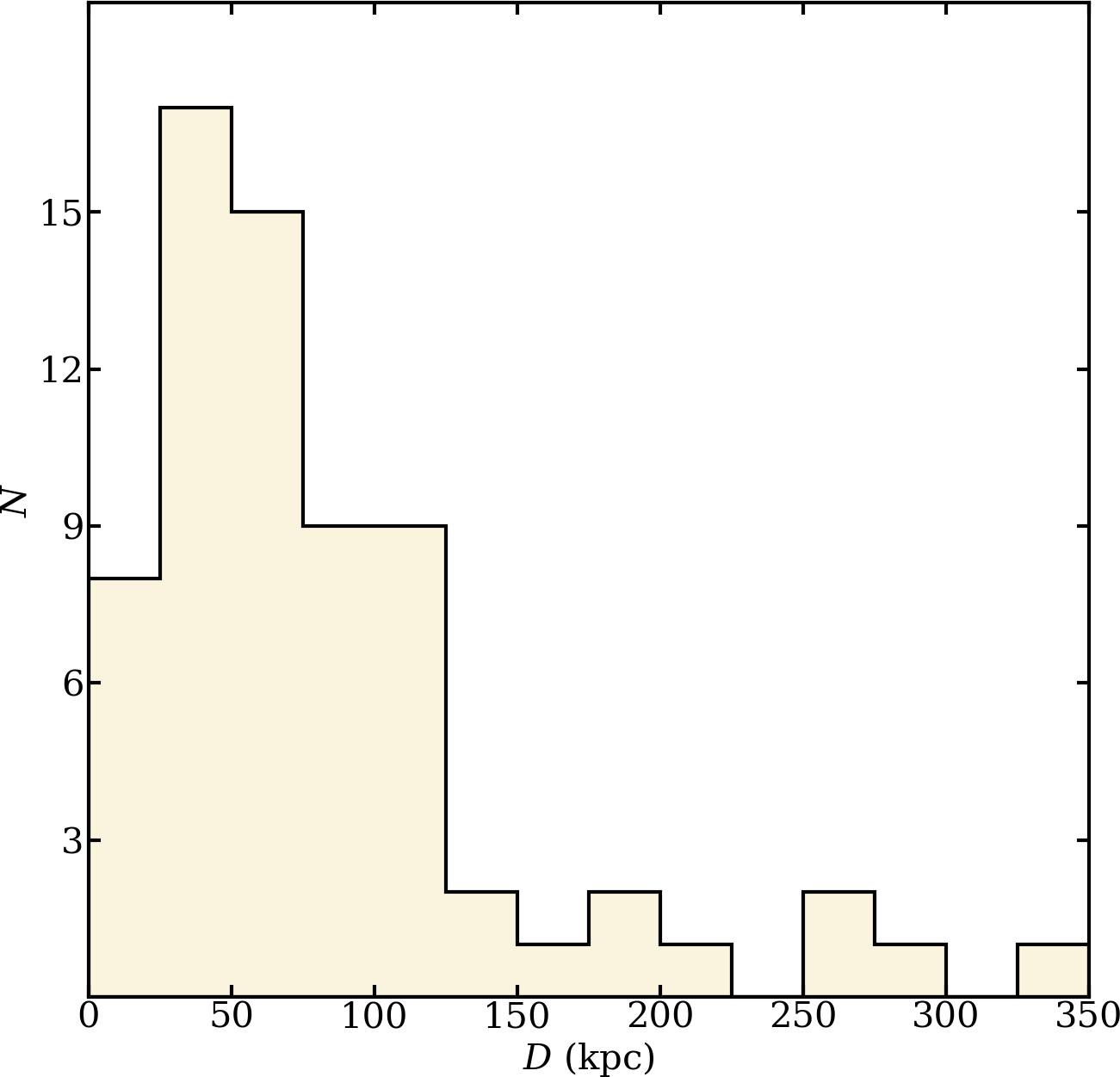}
    \caption{%
    Distribution of spatial extents, $D$, of the resolved BL Lacs.
    \label{fig:core-dominance-against-extent-hist}}
\end{figure}

Fig.~\ref{fig:core-dominance-against-extent-hist} shows the distribution of the spatial extents for the rBZBs, where the median rBZB spatial extent is \SI{69(4)}{kpc}. There is a large spread in the extents, ranging up to \SI{410}{kpc}, and
two sources, %
J1231+3711 ($z = 0.219$), and 
J1340+4410 ($z = 0.546$)
are more than \SI{300}{kpc}.
While this implies large deprojected (although not necessarily unphysical) sizes, it is possible that these jets are bent by some degree (as is typical for astrophysical jets), which could have the effect of reducing the deprojected sizes.

\subsection{Core dominances at 144 MHz} \label{sec:results:core-dominances-at-144-mhz}

\begin{figure}
    \plotone{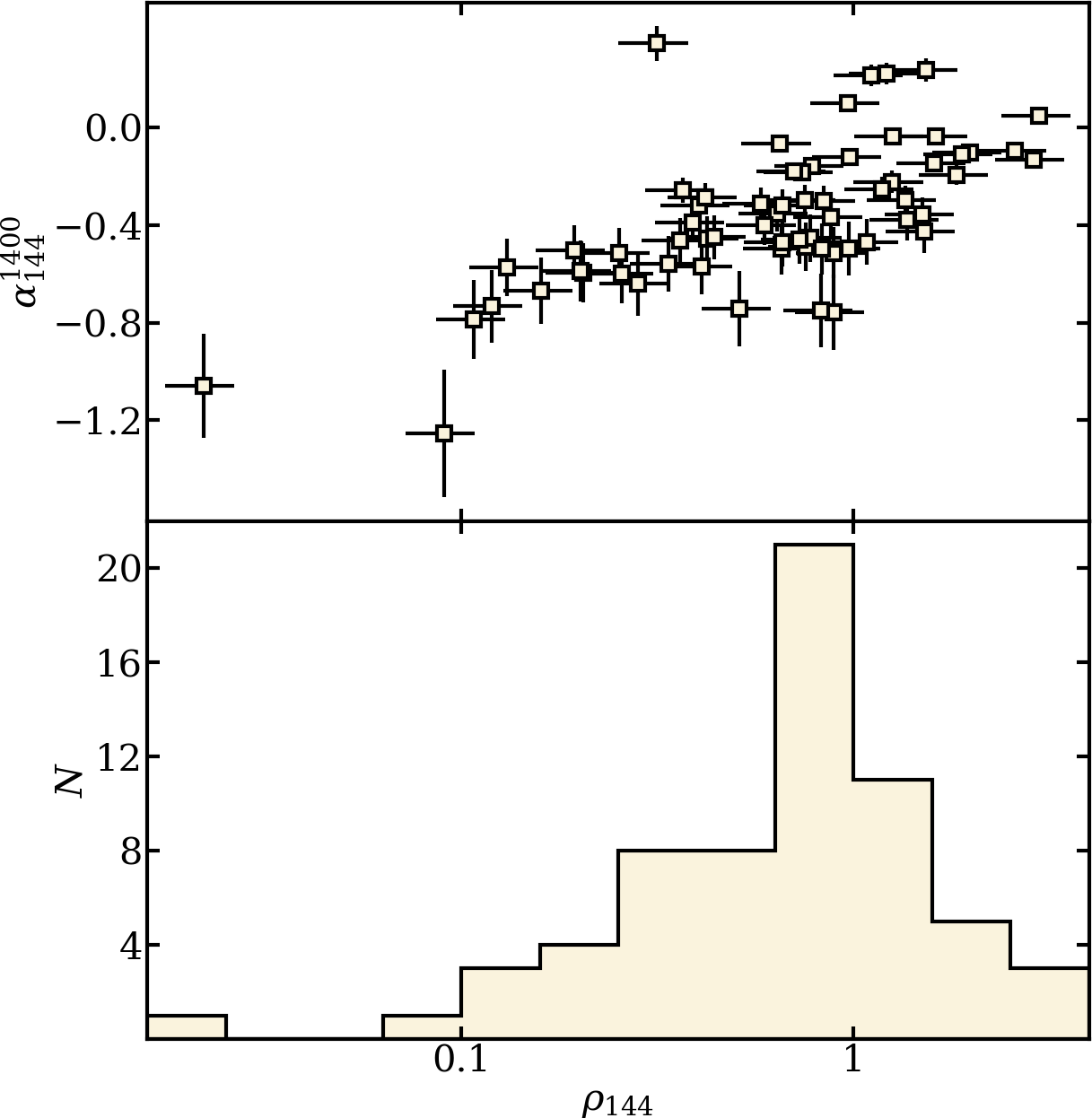}
    \caption{Spectral index, $\alpha_{144}^{1400}$, versus core dominance, $\rho_{144}$, (top) and the distribution of core dominances (bottom) for the resolved BL Lacs.
    \label{fig:spectral-index-against-core-dominance-hist}}
\end{figure}

We find $19/66$ of the rBZBs are core-dominated (i.e. $\rho_{144}\geq 1$). The median core dominance of the rBZBs is $0.74\pm0.06$.
Fig.~\ref{fig:spectral-index-against-core-dominance-hist} (bottom) shows the distribution of $\rho_{144}$ for the rBZBs, which spans two orders of magnitude, reflecting the sensitivity of the beaming factor with respect to the inclination angle and the Lorentz factor.
There is a trend\footnote{The Pearson correlation coefficient, $r$, is reported with the associated probability value, $p$.} ($r=0.63$, $N = 64$, and $p=3\times10^{-8}$) between $\rho_{144}$ and $\alpha^{1400}_{144}$ (Fig.~\ref{fig:spectral-index-against-core-dominance-hist}; top), where the least core-dominated BZBs tend to have steeper $\alpha^{1400}_{144}$.

\subsection{Spectral properties at 144 MHz} \label{sec:results:spectral-properties-at-150-mhz}

\begin{figure}
    \plotone{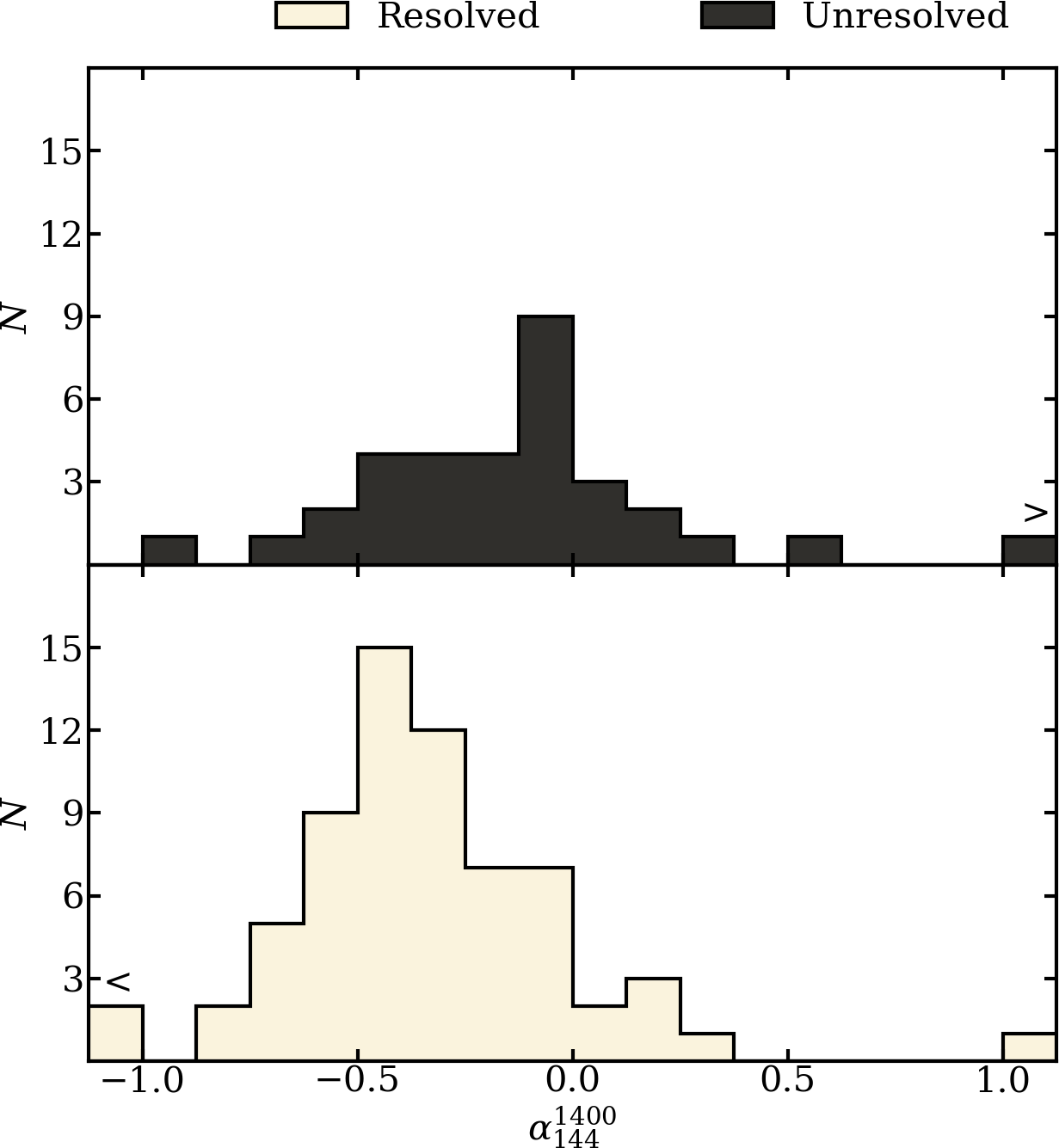}
    \caption{Distribution of \SIrange{144}{1400}{MHz} spectral indices, $\alpha^{1400}_{144}$, of the unresolved (top) and resolved (bottom) BL Lacs.
    \label{fig:extent-against-spectral-index-hist}}
\end{figure}

Of the $99$ sources, $22$ BZBs have steep spectra ($\alpha^{1400}_{144} < -0.5$), $74$ are flat ($-0.5 \leq \alpha^{1400}_{144} \leq 0.5$), and $3$ are inverted ($\alpha^{1400}_{144} > 0.5$). The median \SIrange{144}{1400}{MHz} spectral index for the sample is also flat ($-0.30\pm0.03$). A permutation test \citep{good2013permutation} was used to show that the median spectral index of the well-resolved sources ($\alpha^{1400}_{144} = -0.37\pm0.06$) is steeper than that of the unresolved sources ($\alpha^{1400}_{144} = -0.12\pm0.05$) at a significance level of $p<\num{3e-4}$. The median spectral indices can be classed as flat for both subsamples.
Fig.~\ref{fig:extent-against-spectral-index-hist} shows the distribution of $\alpha^{1400}_{144}$ for the uBZBs (top) and rBZBs (bottom).

\subsection{Radio luminosities and redshifts} \label{sec:results:radio-luminosities}

\begin{figure}
    \plotone{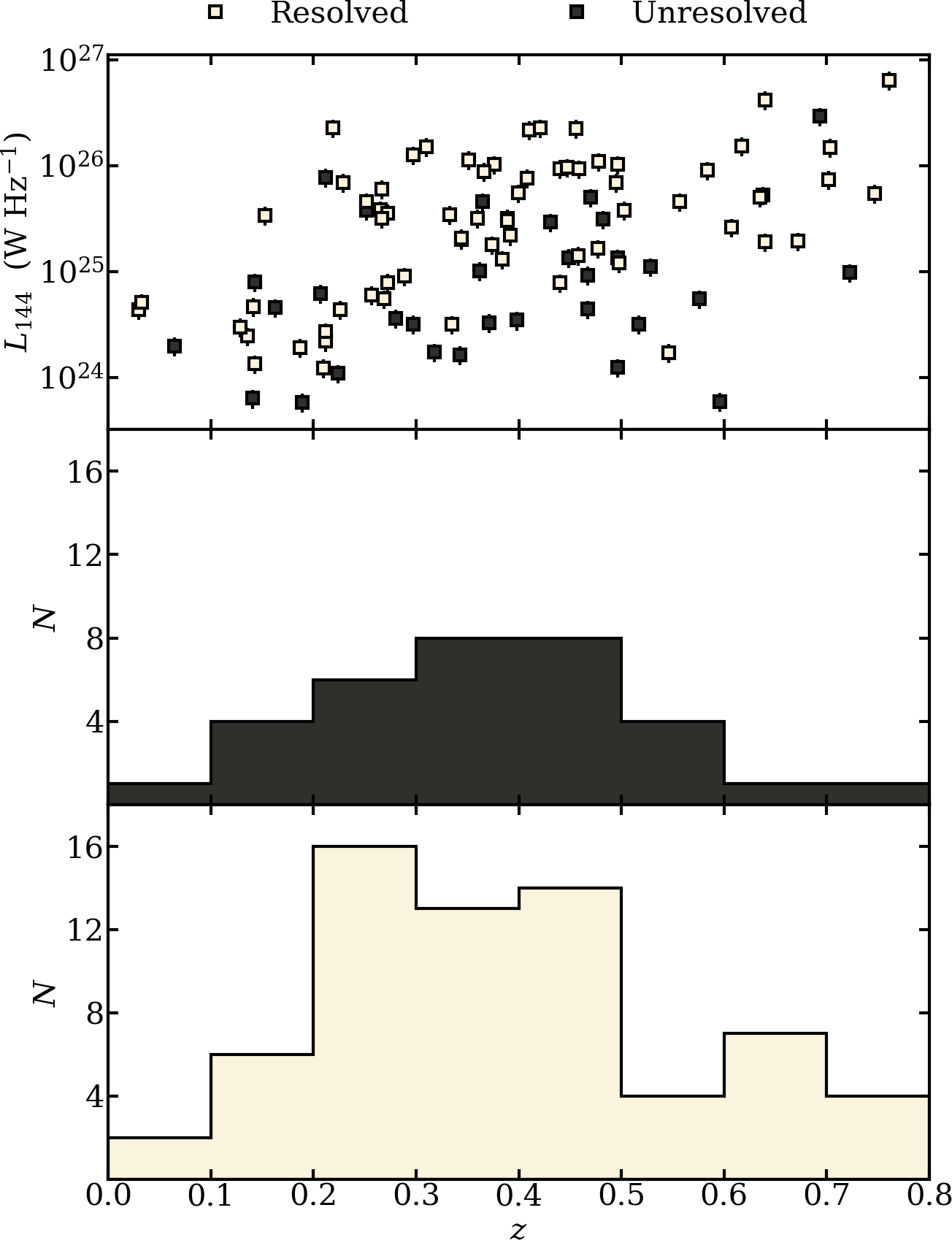}
    \caption{Total luminosity at \SI{144}{MHz}, $L_{144}$, versus redshift, $z$, (top) and the distribution of redshifts for the unresolved (middle) and resolved (bottom) BL Lacs.
    \label{fig:luminosity-against-redshift-hist}}
\end{figure}

The integrated specific luminosities (\si{W\,Hz^{-1}}) of the sources were taking into account using $\alpha^{1400}_{144}$. Fig.~\ref{fig:luminosity-against-redshift-hist} (top) shows the luminosity at \SI{144}{MHz}, $L_{144}$, as a function of redshift, $z$.
While $L_{144}$ loosely increases with $z$, this could be attributed to Malmquist bias \citep{1925MeLuF.106....1M}. 
The redshift distributions for the uBZBs and rBZBs are shown in Fig.~\ref{fig:luminosity-against-redshift-hist} (middle; bottom).
All BZBs in the sample are at $z \leq 0.761$, and the median redshifts of the uBZBs and rBZBs are $0.38\pm0.02$ and $0.37\pm0.04$ respectively. %

\begin{figure}
    \plotone{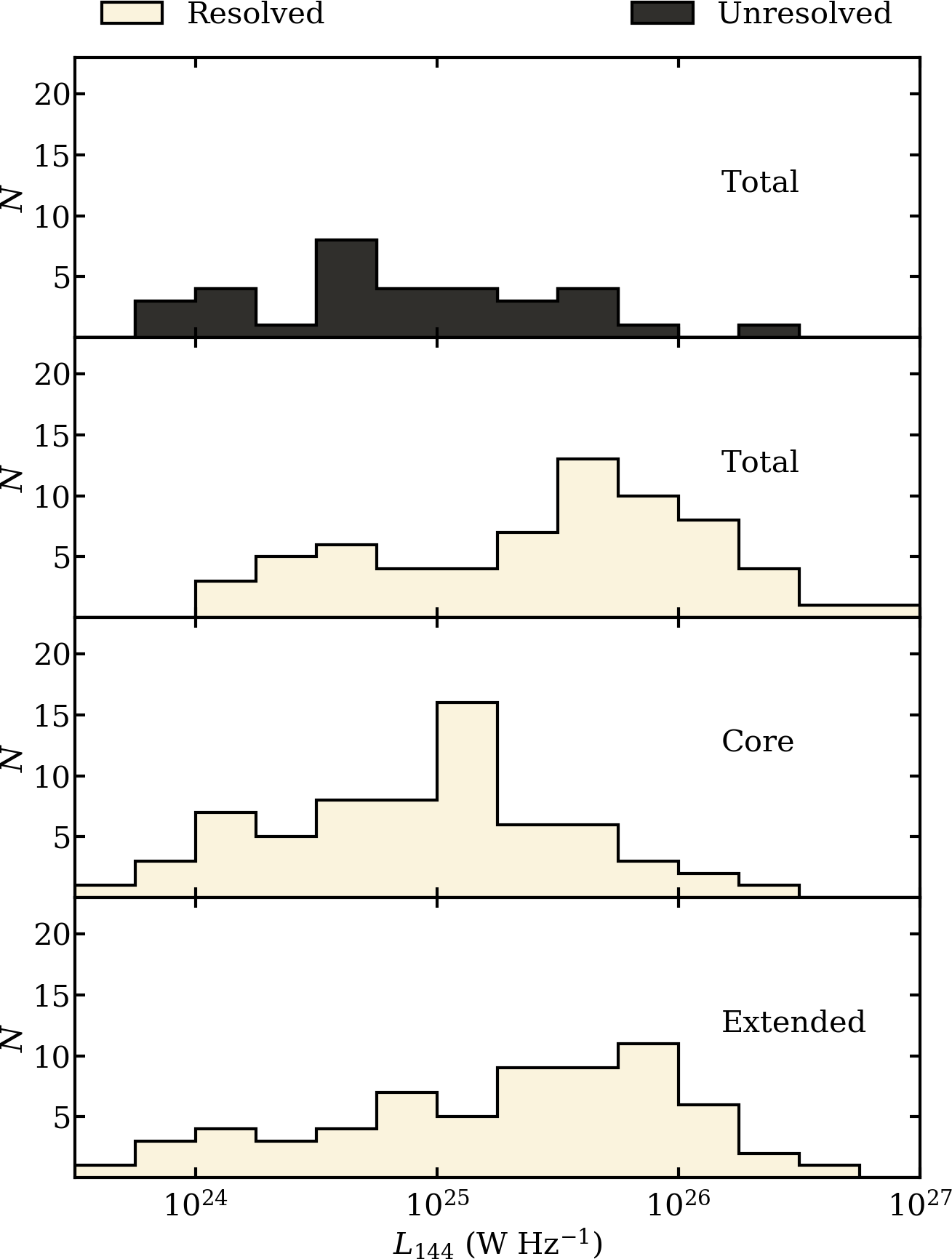}
    \caption{Distribution of (from top to bottom) the \SI{144}{MHz} luminosities for the unresolved BL Lacs, the luminosities for the resolved BL Lacs, the core luminosities for the resolved BL Lacs, and the extended luminosities for the resolved BL Lacs.
    \label{fig:luminosity-hist}}
\end{figure}

The distribution of luminosities for the uBZBs and the rBZBs is shown in Fig.~\ref{fig:luminosity-hist} (top two panels respectively).
There is a clear distinction between these distributions, where rBZBs tend to be more luminous. The rBZBs have higher total luminosities due to the presence of more extended radio emission. %
The luminosities of the core and extended components were then calculated for the rBZBs, where the core and extended spectral indices were assumed to be $0$ and $-0.8$ respectively. This assumption is relatively uncontroversial because the cores of blazars are known to have flat spectral indices, whereas optically-thin synchrotron emission has a spectral index close to $-0.8$.
The distributions for the core and extended luminosities are shown in Fig.~\ref{fig:luminosity-hist} (bottom two panels respectively).
While the rBZBs tend to be more luminous than the uBZBs, the distributions of the core luminosities of the rBZBs and the total luminosities of the uBZBs are similar.
The total and core luminosities are shown as a function of redshift in Table~\ref{tab:my-table}. %

\begin{deluxetable}{@{}ccccccc@{}}
    \tablecaption{Median total and core luminosities of resolved and unresolved BL Lacs. %
        \label{tab:my-table}
    }
    \tablehead{
        \colhead{\multirow{2}{*}{{{\textit{z}}}}} & \multicolumn{3}{c}{{Resolved}} & \colhead{}& \multicolumn{2}{c}{{Unresolved}} \\ \cline{2-4} \cline{6-7}
        \colhead{}  & \colhead{{{\textit{N}}}} & \colhead{{{\textit{L}}\textsubscript{total}}} & \colhead{{\textit{L}\textsubscript{core}}} & \colhead{} & \colhead{{\textit{N}}} & \colhead{{\textit{L}\textsubscript{total}}} 
    }     
    \startdata
    0.0--0.2 & 8  & \num{3.7e24} & \num{1.4e24} &  & 5  & \num{2.0e24}\\
    0.2--0.4 & 29 & \num{3.2e25} & \num{9.2e24} &  & 14 & \num{4.9e24} \\
    0.4--0.6 & 18 & \num{8.4e25} & \num{1.9e25} &  & 12 & \num{1.0e25} \\
    0.6--0.8 & 11  & \num{5.5e25} & \num{2.3e25} & & 2  & \num{1.5e26} \\
    All & 66 & \num{3.5e25} & \num{1.2e25} &  & 33 & \num{6.2e24} \\  
    \enddata
    \tablecomments{Redshift bin is $z$, number of sources is $N$, total luminosity is $L_\mathrm{total}$, and core luminosity is $L_\mathrm{core}$ (both in \si{W\,Hz^{-1}}).}
\end{deluxetable}

\subsection{The radio-to-\texorpdfstring{$\gamma$}{gamma}-ray connection} \label{sec:results:the-gamma-ray-connection}
In total, $53/99$ BZBs had $\gamma$-ray detections; properties of the $\gamma$-ray BZBs are given in Table~\ref{tab:gamsources}. 
The medians of the logarithms of the integrated photon fluxes at \SIrange{1}{100}{\giga\electronvolt} for the rBZBs ($-9.61\pm0.08$) and uBZBs ($-9.61\pm0.13$) are the same within uncertainty.
The $\gamma$-ray-detected and non-$\gamma$-ray-detected BZBs are similar in terms of spatial extents, angular extents, luminosities and core dominances. However, the $\gamma$-ray BZBs tend to have smaller redshifts. The median radio spectral index of the $\gamma$-ray BZBs is also slightly flatter than that of the BZBs without a detection.

\begin{figure}
    \plotone{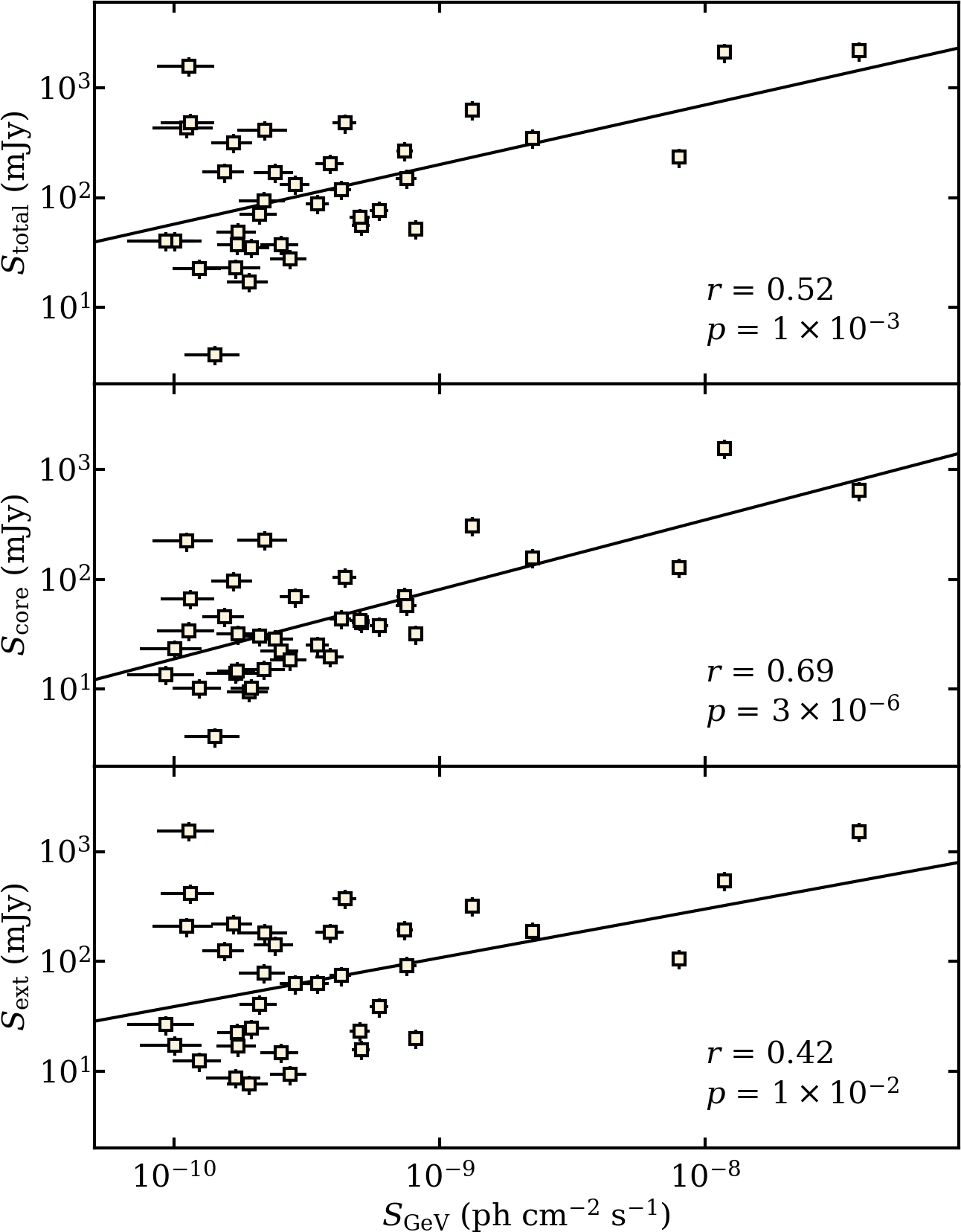}
    \caption{Total flux density, $S_\mathrm{total}$, (top), core flux density, $S_\mathrm{core}$ (middle), and extended flux density, $S_\mathrm{ext}$ (bottom) at \SI{144}{MHz} versus $\gamma$-ray photon flux, $S_\mathrm{GeV}$, for the $\gamma$-ray-detected resolved BL Lacs. The correlation coefficient and statistical significance are $r$ and $p$ respectively.
    \label{fig:gamma-ray-connection}}
\end{figure}

Fig.~\ref{fig:gamma-ray-connection} shows the total (top), core (middle), and extended (bottom) \SI{144}{MHz} flux density against $S_\mathrm{GeV}$ for the rBZBs.
The Pearson correlation coefficient between $S_\mathrm{total}$ and $S_\mathrm{GeV}$ is $0.52$ ($N = 36$ and $p = \num{1e-3}$).
For $S_\mathrm{core}$ and $S_\mathrm{GeV}$, $r$ increases to $0.69$ ($N = 36$ and $p = \num{3e-6}$), while for $S_\mathrm{ext}$ and $S_\mathrm{GeV}$, $r = 0.42$ ($N = 35$ and $p = \num{1e-2}$).

\begin{deluxetable}{lhhhhhhRRh}
    \tablecaption{
        Median properties of the $\gamma$-ray and non-$\gamma$-ray detected BL Lacs. Reported uncertainties are the standard error. %
        \label{tab:gamsources}
    }
    \tablehead{
        \colhead{} & & & & & & &\colhead{$\gamma$-ray} & \colhead{Non-$\gamma$-ray} & \nocolhead{Total} %
    }     
    \startdata
    $N$&$ 34 $ & $ 27 $&$ 61 $&$ 18 $&$ 20 $&$ 38 $&$ 53 $&$ 46 $&$ 99 $ \\
    $z$&$ 0.29 \pm 0.03 $&$ 0.47 \pm 0.03 $&$ 0.38 \pm 0.02 $&$ 0.29 \pm 0.03 $&$ 0.48 \pm 0.03 $&$ 0.39 \pm 0.03 $&$ 0.29 \pm 0.04 $&$ 0.46 \pm 0.02 $&$ 0.38 \pm 0.02 $\\
    $\log_{10} L_{144}$ &$ 5 \pm 1 $&$ 9 \pm 3 $&$ 7 \pm 1 $&$ 3 \pm 2 $&$ 3 \pm 1 $&$ 3 \pm 1 $&$ 25.3 \pm 0.3 $&$ 25.3 \pm 0.1 $&$ 5 \pm 1 $\\
    $\alpha^{1400}_{144}$&$ -0.22 \pm 0.07 $&$ -0.43 \pm 0.05 $&$ -0.31 \pm 0.05 $&$ -0.14 \pm 0.05 $&$ -0.2 \pm 0.1 $&$ -0.15 \pm 0.07 $&$ -0.26 \pm 0.05 $&$ -0.38 \pm 0.06 $&$ -0.25 \pm 0.04 $\\
    $\log_{10}S_\mathrm{GeV}$ &$ 2.3\times10^{10} $&\nodata&$ \nodata $&$ 0.5 \pm 0.2 $&\nodata&\nodata&$ -9.6\pm0.1 $&\nodata&\nodata\\
    $N_\mathrm{ext}$&$ 32 \pm 4 $&$ 41 \pm 7 $&$ 32 \pm 2 $&\nodata&\nodata&\nodata&$36$&$30$&\nodata\\
    $\rho_{144}$&$ 1.7 \pm 0.3 $&$ 1.1 \pm 0.3 $&$ 1.4 \pm 0.2 $&\nodata&\nodata&\nodata&$ 0.83 \pm 0.17 $&$ 0.72 \pm 0.05 $&\nodata\\
    $D$ (kpc)&$ 201 \pm 23 $&$ 234 \pm 25 $&$ 215 \pm 17 $&\nodata&\nodata&\nodata&$70\pm15$ &$ 69 \pm 6 $&\nodata\\
    $\Phi$ (\si{\arcsecond})&$ 27 \pm 5 $&$ 41 \pm 7 $&$ 32 \pm 2 $&\nodata&\nodata&\nodata&$ 15 \pm 3 $&$ 12 \pm 1 $&\nodata\\
    \enddata
    \tablecomments{Number of sources is $N$, redshift is $z$, \SI{144}{MHz} luminosity is $L_{144}$, \SIrange{144}{1400}{MHz} spectral index is $\alpha^{1400}_{144}$, $\gamma$-ray flux is $S_\mathrm{GeV}$, %
    number of extended BL Lacs is $N_\mathrm{ext}$, \SI{144}{MHz} core dominance of extended BL Lacs is $\rho_{144}$, spatial extent of extended BL Lacs is $D$, and angular extent of extended BL Lacs is $\Phi$.} %
\end{deluxetable}

\section{Discussion} \label{sec:discussion}

Previous studies identified extended emission at \si{GHz} frequencies around BZBs \citep[e.g.][]{1983ApJ...266...18U}, but the preliminary LoTSS Second Data Release has made it possible to systematically study the morphology of BZBs at \SI{144}{MHz} for the first time, where we expect to find a larger fraction of diffuse emission due to the steeper spectra of the extended structures.
We identified extended emission around $66/99$ BZBs at \SI{144}{MHz}. 
We compare BZBs and LERGs because radio-loud AGN unification theories predict that these observational source classifications are the same source type intrinsically \citep{2020ApJ...900L..34M,2020ApJS..247...71M,2020A&A...633A.161C}.
The spatial extents spanned a wide range (up to \SI{410}{kpc}), and this is in line with predictions of the AGN unification scenario, where BZBs are LERGs viewed end-on, because LERGs %
range from several \si{kpc} to several \si{Mpc} in the plane of the sky.
For the BZBs that were not found to be extended, it is possible that the resolution is insufficient to spatially resolve these sources, or that extended low-surface brightness radio emission exists below the LDR2 sensitivity.
Particularly for the BZBs with $S_\mathrm{total} \lesssim \SI{40}{mJy}$, the extended emission may be below the detection threshold of LDR2.

The core luminosities per redshift bin of the rBZBs are comparable to the total luminosities of the uBZBs, suggesting no major difference in intrinsic core power between the two sources. That is, at a given core power we find large ranges of $D$ and $L_\mathrm{ext}$. A similar effect has also been seen in Faranoff-Riley class 0 sources {\citep[FR-0s;][]{2015A&A...576A..38B} and FR-Is, where
\cite{2020A&A...633A.161C} found that
for a given core power, FR-0s tend to have fainter diffuse emission than FR-Is.
The key variables controlling whether a BZB is extended or not is unclear, and could relate to the initial conditions or the environment. 
It is possible that the uBZBs have either only switched on recently, where the AGN activity is too recent to have produced plumes, or perhaps they have been switched off for $\gtrsim 10^8$ years, as suggested by \citet{2015ApJ...812...79P}, where this timescale is an upper limit based on the fact that the synchrotron break frequency drops to $\lesssim \SI{100}{MHz}$, assuming $\alpha_{144}^{1400} \approx -0.8$, a $\sim\SI{1}{\micro G}$ magnetic field, and no new injections of energetic particles \citep{1986MNRAS.219..575C,2007MNRAS.378..581J}. %

We directly measured the core dominances of the extended BZBs by separating the core and extended flux components.
Our results are broadly in line with \citet{2016A&A...588A.141G} and \citet{2018ApJ...869..133F}, where the core dominances of blazars were calculated by decomposing $S_\mathrm{total}$ using estimates of the core and extended spectral indices.
While the core component contributes more than half of the total flux for $19/66$ of the rBZBs, the majority of these extended BZBs were not core-dominated.
This finding is in agreement with \citet{2019MNRAS.490.5798D} where they state that deeper low frequency surveys should result in lower core dominances because more radio lobe emission can be detected.
Thus, the core dominance may not be a reliable parameter for selecting new blazar candidates in future low frequency surveys.

The median radio spectral index for the sample is flat, which confirms several recent studies that have found that BZBs, and blazars generally, have flat spectral indices down to \SI{\sim100}{\mega\hertz} at least \citep{2016A&A...588A.141G,2019A&A...622A..14M,2019MNRAS.490.5798D}. However, the most extended BZBs tended to have spectral indices closer to $-0.8$, similar to classic radio galaxies, likely due to the presence of diffuse emission.

The population of relativistic electrons that give rise to the beamed radio emission are thought to be responsible for producing the $\gamma$-ray emission, by upscattering seed photons from external radiation fields to $\gamma$-ray energies.
This is believed to be also the mechanism behind the radio-to-$\gamma$-ray connection at GHz frequencies.
Previous studies have shown that this correlation weakens with decreasing radio frequency \citep[e.g.][]{2016A&A...588A.141G,2019A&A...622A..14M}.
By calculating the radio-to-$\gamma$-ray correlation using the core and extended flux densities at \SI{144}{MHz} separately, we have shown that this connection weakens at low radio frequencies because there is an increase in diffuse emission.
The diffuse MHz emission is not expected to correlate strongly with the $\gamma$-ray emission because the diffuse MHz emission is unbeamed, and the electron population producing the diffuse MHz emission is likely to be distinct from the electrons that give rise to the $\gamma$-rays, reflecting the time-integrated history of jet activity rather than the instantaneous view given by the core emission.
In contrast, the beamed core emission at low frequencies still correlates reasonably well with the $\gamma$-ray photon flux.

\section{Conclusions and future prospects} \label{sec:summary-and-conclusions}

We have presented the first morphological study of a sample of BZBs at \SI{144}{MHz}. Our findings are as follows:

\begin{itemize}
    \item Extended emission was revealed around $66/99$ BZBs. %
    The distributions of spatial extents and the extended component luminosities at \SI{144}{MHz} are consistent with expectations based on the AGN unification paradigm, where BZBs are the aligned counterparts of LERGs. For a given core luminosity and redshift, a range of spatial extents were found.
    \item The median integrated \SIrange{144}{1400}{MHz} spectral index for the BZBs was $-0.30\,\pm\,0.03$, confirming that the low frequency spectra of BZBs are still dominated by the beamed core component, but this depends on the spatial extent, with the spectral index being steeper for more extended BZBs.
    
    \item Of the $66$ rBZBs, $19$ were core-dominated at \SI{144}{MHz}, and the most core-dominated sources tended to have smaller spatial extents and flatter spectral indices.
    
    \item Positive correlations were identified between the \SIrange{1}{100}{GeV} flux and both the \SI{144}{MHz} total ($r=0.52$) and core ($r=0.69$) flux densities for the rBZBs. The correlation between the $\gamma$-ray flux and the extended emission was weaker ($r=0.42$). %
    This suggests that the \SI{144}{MHz} emission from the core is more likely to be directly related to the present nuclear activity than the large-scale extended emission.
\end{itemize}

Characterising the diffuse emission around BZBs can help to explain some of the outstanding questions regarding blazars. For example, measurements of the diffuse flux allow for estimates of the jet power via scaling relations \citep{2011ApJ...740...98M}.
Calculating the jet power can shed light on the relationship between the jet power and the AGN accretion rate, while estimates of the diffuse flux and the spatial extent can also be used to characterise the environments.
We focused on BZBs using LoTSS, the most sensitive low frequency survey in existence. %
In future, LoTSS in-band spectral index maps will make it possible to distinguish between the beamed jet flux and the unbeamed diffuse flux that contribute to the extended emission.
Other forthcoming surveys will help further constrain the \SI{144}{MHz} morphology of BZBs.
The LOFAR LBA Sky Survey (LoLSS; de Gasperin et al., in prep) will eventually cover the northern hemisphere sky at \SIrange{42}{66}{MHz} at a sensitivity of \SI{1}{mJy\,beam^{-1}} and \SI{15}{\arcsecond} resolution.
With LoLSS, we will be able to identify possible ultra-steep spectrum radio halos around the unresolved BZBs that are undetectable in LDR2.
The future sky survey with the LOFAR international stations (Morabito et al. in prep) will image the \SIrange{122}{168}{MHz} sky at \SI{0.3}{\arcsecond} resolution, potentially spatially resolving some of the unresolved BZBs.
In addition, optical spectra provided by the WEAVE-LOFAR survey \citep{2016sf2a.conf..271S} will be beneficial in identifying LDR2 counterparts to unidentified $\gamma$-ray sources.
In the long term, surveys with the Square Kilometre Array (SKA) will also be key in studying this population \citep{Kharb2016}.

\begin{longrotatetable}
    \begin{deluxetable*}{
        c  %
        C  %
        C  %
        C  %
        c  %
        C  %
        C  %
        C  %
        C  %
        C  %
        C  %
        C  %
    }
    \tablecaption{
        Details of the BZB sample.
        \label{tab:results}
    }
    \tabletypesize{\footnotesize}
    \tablehead{
        \colhead{Name} & \colhead{RA} & \colhead{Dec} & \colhead{$z$} & \colhead{Resolved?} & \colhead{$S_\mathrm{total}$} & \colhead{$S_\mathrm{GeV}$} & \colhead{$\alpha_{144}^{1400}$} & \colhead{$\rho_{144}$} & \colhead{$D$} &  \colhead{$\Phi$} & \colhead{$L_{144}$} \\
        \colhead{} & \colhead{\si{\degree}} & \colhead{\si{\degree}} & \colhead{} & \colhead{} & \colhead{\si{\milli Jy}} &  \colhead{\si{ph \, \centi \metre^{-2} \second^{-1}}} & \colhead{} & \colhead{} & \colhead{\si{\kilo pc}} & \colhead{\si{\arcsecond}} & \colhead{\si{\watt \per \hertz}}
    }
    \decimals
    \colnumbers
    \startdata
J0047+3948&11.9801&39.8160&0.252 & T & $ 264 \pm 53 $  &                  $ (7.4 \pm 0.5) \times 10^{-10} $ &                                 $ -0.46 \pm 0.09 $&                    $ 0.36 \pm 0.07 $    &                     $ 68.1\pm0.4 $     &                     $ 17.1\pm0.1$             &                           $ (4.6 \pm 0.9) \times 10^{25} $\\
J0112+2244&          18.0243        &                22.7441     &                     0.265   &        T        &         $ 233 \pm 47 $  &                  $ (8.0 \pm 0.2) \times 10^{-9} $  &                                 $ 0.22 \pm 0.04 $ &                    $ 1.21 \pm 0.25 $    &                     $ 38.3\pm0.1 $     &                     $ 9.3\pm0.1 $             &                           $ (3.9 \pm 0.8) \times 10^{25} $\\
J0123+3420&          20.7860&                34.3468     &                     0.272   &        T        &         $ 169 \pm 34 $  &                  $ (2.4 \pm 0.4) \times 10^{-10} $ &                                 $ -0.60 \pm 0.12 $&                    $ 0.20 \pm 0.04 $  &                     $ 95.9\pm0.4 $     &                     $ 22.9\pm0.1 $             &                           $ (3.6 \pm 0.7) \times 10^{25} $\\
J0203+3042&          30.9345&                30.7105              &                     0.761   &        T        &         $ 345 \pm 69 $  &  $ (2.24 \pm 0.09) \times 10^{-9} $ &                                 $ -0.30 \pm 0.06 $&                    $ 0.84 \pm 0.17 $    &                     $ 130.7\pm0.3 $     &                     $ 17.4\pm0.1 $             &                           $ (6.4 \pm 1.3) \times 10^{26} $\\
    $\vdots$ & $\vdots$ & $\vdots$ & $\vdots$ & $\vdots$ & $\vdots$ & $\vdots$ & $\vdots$ & $\vdots$ & $\vdots$ & $\vdots$ & $\vdots$ \\
J1808+3520$\dagger$&          272.2071&                35.3452     &                     0.141  &        T        &         $ 94 \pm 19 $   &                  $ (2.2 \pm 0.4) \times 10^{-10} $ &                                 $ -0.50 \pm 0.10 $&                    $ 0.19 \pm 0.04 $  &                     $ 72.7\pm0.8$     &                     $ 29.1\pm0.3 $             &                           $ (4.7 \pm 0.9) \times 10^{24} $\\
J2229+2255$\dagger$&          337.2966&                22.9166     &                     0.440    &        T        &         $ 18 \pm 4 $    &                  \nodata                          &                 $0.35\pm0.07$                 &                    $ 0.31 \pm 0.06 $    &                     $ 58.3\pm1.7 $     &                     $ 10.1\pm0.3 $             &                           $ (8.9 \pm 1.8) \times 10^{24} $\\
J2237+1840$\dagger$&          339.2701 &                18.6822     &                     0.722  &        F        &         $ 7 \pm 1 $     &                  \nodata                          &                                 $ -0.08 \pm 0.02 $&                    \nodata    &                     $ \leq 28.1\pm0.7 $ &                     $ \leq 3.8\pm0.1 $         &                           $ (9.9 \pm 2.0) \times 10^{24} $\\
J2343+3439&          355.8899&                34.6641     &                     0.366   &        T        &         $ 203 \pm 41 $  &                  $ (3.9 \pm 0.5) \times 10^{-10} $ &                                 $ -0.79 \pm 0.16 $&                    $ 0.11 \pm 0.02 $  &                     $ 151.9\pm1.7 $     &                     $ 29.6\pm0.3 $             &                           $ (8.9 \pm 1.8) \times 10^{25} $\\    
    \enddata
    \tablecomments{(1) Source name. (2--3) Coordinates in the J2000 frame. (4) Redshift. (5) A flag stating whether the source is resolved (`T') or not (`F'). (6) Integrated flux density at \SI{144}{\mega\hertz} from LDR2. (7) $\gamma$-ray flux from 4LAC. (8) \SIrange{144}{1400}{\mega\hertz} spectral index with LDR2 and NVSS. (9) Core dominance. (10) Spatial extent. (11) Angular extent. (12) Radio luminosity. \\
    This table is available in its entirety in machine-readable form.\\
    $\dagger$ BZBs not in Roma-BZCAT v5.0.}
    \end{deluxetable*}
\end{longrotatetable}

\acknowledgments

SM acknowledges support from the Irish Research Council Postgraduate Scholarship and UCD U21 funding.
BM acknowledges support from the UK Science and Technology Facilities Council (STFC) under grants ST/R00109X/1 and ST/R000794/1. 
MJH acknowledges support from STFC [ST/R000905/1].
FM acknowledges financial contribution from the agreement ASI-INAF n.2017-14-H.0.
This work is supported by the ``Departments of Excellence 2018--2022'' grant awarded by the Italian Ministry of Education, University and Research (MIUR) (L. 232/2016).
This research has made use of resources provided by the Compagnia di San Paolo for the grant awarded on the BLENV project (S1618\_L1\_MASF\_01) and by the Ministry of Education, Universities and Research for the grant MASF\_FFABR\_17\_01.
This investigation is supported by the National Aeronautics and Space Administration (NASA) grants GO4-15096X, AR6-17012X, GO6-17081X, GO9-20083X, and GO0-21110X.

LOFAR is the Low Frequency Array designed and constructed by ASTRON. It has observing, data processing, and data storage facilities in several countries, which are owned by various parties (each with their own funding sources), and which are collectively operated by the ILT foundation under a joint scientific policy. The ILT resources have benefited from the following recent major funding sources: CNRS-INSU, Observatoire de Paris and Universit\'e d'Orl\'eans, France; BMBF, MIWF-NRW, MPG, Germany; Science Foundation Ireland (SFI), Department of Business, Enterprise and Innovation (DBEI), Ireland; NWO, The Netherlands; The Science and Technology Facilities Council, UK; Ministry of Science and Higher Education, Poland; The Istituto Nazionale di Astrofisica (INAF), Italy.

This research made use of the Dutch national e-infrastructure with support of the SURF Cooperative (e-infra 180169) and the LOFAR e-infra group. The J\"ulich LOFAR Long Term Archive and the German LOFAR network are both coordinated and operated by the J\"ulich Supercomputing Centre (JSC), and computing resources on the supercomputer JUWELS at JSC were provided by the Gauss Centre for Supercomputing e.V. (grant CHTB00) through the John von Neumann Institute for Computing (NIC).

This research made use of the University of Hertfordshire high-performance computing facility and the LOFAR-UK computing facility located at the University of Hertfordshire and supported by STFC [ST/P000096/1], and of the Italian LOFAR IT computing infrastructure supported and operated by INAF, and by the Physics Department of Turin university (under an agreement with Consorzio Interuniversitario per la Fisica Spaziale) at the C3S Supercomputing Centre, Italy.

The Pan-STARRS1 Surveys (PS1) and the PS1 public science archive have been made possible through contributions by the Institute for Astronomy, the University of Hawaii, the Pan-STARRS Project Office, the Max-Planck Society and its participating institutes, the Max Planck Institute for Astronomy, Heidelberg and the Max Planck Institute for Extraterrestrial Physics, Garching, The Johns Hopkins University, Durham University, the University of Edinburgh, the Queen's University Belfast, the Harvard-Smithsonian Center for Astrophysics, the Las Cumbres Observatory Global Telescope Network Incorporated, the National Central University of Taiwan, the Space Telescope Science Institute, the National Aeronautics and Space Administration under Grant No. NNX08AR22G issued through the Planetary Science Division of the NASA Science Mission Directorate, the National Science Foundation Grant No. AST-1238877, the University of Maryland, Eotvos Lorand University (ELTE), the Los Alamos National Laboratory, and the Gordon and Betty Moore Foundation.

The National Radio Astronomy Observatory is a facility of the National Science Foundation operated under cooperative agreement by Associated Universities, Inc.  %

\vspace{10mm}  %
\facilities{LOFAR, Haleakala Observatory (Pan-STARRS), and VLA.}

\software{DDFacet \citep{2018A&A...611A..87T},
          KillMS \citep{2014A&A...566A.127T, 2015MNRAS.449.2668S},
          Python \citep{CS-R9526},
          NDPPP \citep{2018ascl.soft04003V},
          NumPy \citep{2011CSE....13b..22V},
          Astropy \citep{2013A&A...558A..33A},
          TOPCAT \citep{2013StaUN.253.....T},
          PyBDSF \citep{2015ascl.soft02007M}, and
          Matplotlib \citep{Hunter:2007}.
          }

\bibliography{references}{}
\bibliographystyle{aasjournal}
\end{document}